\documentclass[twocolumn,superscriptaddress,showpacs,preprintnumbers,showkeys,amsmath,amssymb,aps,prl,floatfix]{revtex4-2}
\usepackage{amsmath}
\usepackage{xcolor}
\usepackage{graphicx}
\usepackage{bm}
\usepackage[colorlinks=true, citecolor=blue, urlcolor=blue, linkcolor=blue]{hyperref}
\usepackage[normalem]{ulem}
\usepackage{soul}
\usepackage{gensymb}
\usepackage{siunitx}

\newcommand{\NaCCOC}{Na$_x$Ca$_{2-x}$CuO$_2$Cl$_2$}

\newcommand{\p}{^{\prime}}
\newcommand{\pp}{^{\prime\prime}}

\begin{document}

\title{On the cuprates' universal waterfall feature: evidence of a momentum-driven crossover}

\author{Benjamin Bacq-Labreuil}
\email{benjamin.bacq-labreuil@ipcms.unistra.fr}
\affiliation{CPHT,  CNRS,  Ecole  Polytechnique,  Institut  Polytechnique  de  Paris,  F-91128  Palaiseau,  France}
\affiliation{D\'epartement de physique, Regroupement qu\'eb\'ecois sur les mat\'eriaux de pointe \& Institut quantique Universit\'e de Sherbrooke, 2500 Boul. Universit\'e, Sherbrooke, Qu\'ebec J1K2R1, Canada}

\author{Chafic Fawaz}
\affiliation{Univ. Grenoble Alpes, CNRS, Grenoble INP, Institut Néel, F-38000 Grenoble, France}

\author{Yuichi Okazaki}
\affiliation{Department of Materials Science, Graduate School of Engineering
Osaka Metropolitan University,
1-1 Gakuen-cho, Naka-ku, Sakai, Osaka 599-8531, Japan}

\author{Yukiko Obata}
\altaffiliation[Present address: ]{Ultra-low Temperature Physics Laboratory, Faculty of Mathematics and Physics, Institute of Science and Engineering,  Kanazawa University, Kakumamachi, Kanazawa 920-1192, Japan}
\affiliation{Tokyo Tech World Research Hub Initiative (WRHI), Institute of Innovative Research, Tokyo Institute of Technology, 4259 Nagatsuta, Midori-ku, Yokohama, Kanagawa 226-8503, Japan}

\author{Hervé Cercellier}
\affiliation{Univ. Grenoble Alpes, CNRS, Grenoble INP, Institut Néel, F-38000 Grenoble, France}

\author{Patrick Lefevre}
\affiliation{Synchrotron SOLEIL, L’Orme des Merisiers, Départementale 128, 91190 Saint-Aubin, France}

\author{Fran\c{c}ois Bertran}
\affiliation{Synchrotron SOLEIL, L’Orme des Merisiers, Départementale 128, 91190 Saint-Aubin, France}

\author{David Santos-Cottin}
\affiliation{Department of Physics, University of Fribourg, 1700 Fribourg, Switzerland}

\author{Hajime Yamamoto}
\altaffiliation[Permanent address: ]{Institute of Multidisciplinary Research for Advanced Materials (IMRAM), Tohoku University Katahira 2-1-1, Aoba-ku, Sendai 980-8577, Japan}
\affiliation{Materials and Structures Laboratory, Institute of Integrated Research, Institute of Science Tokyo, 4259 Nagatsuta, Midori-ku, Yokohama, 226-8503, Japan}

\author{Ikuya Yamada}
\affiliation{Department of Materials Science, Graduate School of Engineering
Osaka Metropolitan University,
1-1 Gakuen-cho, Naka-ku, Sakai, Osaka 599-8531, Japan}

\author{Masaki Azuma}
\affiliation{Materials and Structures Laboratory, Institute of Integrated Research, Institute of Science Tokyo, 4259 Nagatsuta, Midori-ku, Yokohama, 226-8503, Japan}
\affiliation{Kanagawa Institute of Industrial Science and Technology, Ebina 243-0435, Japan}

\author{Koji Horiba} 
\affiliation{Photon Factory and Condensed Matter Research Center, Institute of Materials Structure Science, High Energy Accelerator Research Organization (KEK), Tsukuba 305-0801, Japan.}

\author{Hiroshi Kumigashira} 
\affiliation{Photon Factory and Condensed Matter Research Center, Institute of Materials Structure Science, High Energy Accelerator Research Organization (KEK), Tsukuba 305-0801, Japan.}
\affiliation{Institute of Multidisciplinary Research for Advanced Materials (IMRAM), Tohoku University Katahira 2-1-1, Aoba-ku, Sendai 980-8577, Japan}

\author{Matteo d'Astuto}
\email{matteo.dastuto@neel.cnrs.fr}
\affiliation{Univ. Grenoble Alpes, CNRS, Grenoble INP, Institut Néel, F-38000 Grenoble, France}
\affiliation{Tokyo Tech World Research Hub Initiative (WRHI), Institute of Innovative Research, Tokyo Institute of Technology, 4259 Nagatsuta, Midori-ku, Yokohama, Kanagawa 226-8503, Japan}

\author{Silke Biermann}
\email{silke.biermann@cpht.polytechnique.fr}
\affiliation{CPHT,  CNRS,  Ecole  Polytechnique,  Institut  Polytechnique  de  Paris,  F-91128  Palaiseau,  France}
\affiliation{Coll{\`e}ge  de  France,  11  place  Marcelin  Berthelot,  F-75005  Paris,  France}
\affiliation{Department  of  Physics,  Division  of  Mathematical  Physics,Lund  University,  Professorsgatan  1,  22363  Lund,  Sweden}
\affiliation{European  Theoretical  Spectroscopy  Facility, F-91128 Palaiseau, Europe}

\author{Benjamin Lenz}
\email{benjamin.lenz@upmc.fr}
\affiliation{IMPMC, Sorbonne Universit\'e, CNRS, MNHN, 4 place Jussieu, F-75252 Paris, France}

\date{\today}

\begin{abstract}

We study two related universal anomalies of the spectral function of cuprates, so called waterfall and high-energy kink features, by a combined cellular dynamical mean-field theory and angle-resolved photoemission study for the oxychloride \NaCCOC~(Na-CCOC).
Tracing their origin back to an interplay of spin-polaron and local correlation effects both in undoped and hole-doped (Na-)CCOC, we establish them as a universal crossover between regions differing in the momentum dependence of the coupling and not necessarily in the related quasiparticles' energies. 
The proposed scenario extends to doping levels coinciding with the cuprate's superconducting dome and motivates further investigations of the fate of spin-polarons in the superconducting phase.

\end{abstract}

\maketitle


Understanding the physics of cuprate high-temperature superconductors (HTSC) remains one of the most intricate challenges of condensed matter physics. 
Among the tools available to tackle this long-standing problem, photoemission spectroscopy is a method of choice for it provides a detailed access to the materials' electronic structure.
Two related universal spectral features of the cuprate family have particularly attracted the attention: the so-called \emph{waterfall} and \emph{high-energy kink} features~\cite{damascelli2003,graf2007}. 
They have been detected in angle-resolved photoemission spectroscopy (ARPES) measurements performed on hole-~\cite{kohsaka2003,graf2007a,graf2007,kordyuk2006,borisenko2006a,borisenko2006b,valla2007,xie2007,moritz2009,zhang2008,damascelli2003,basak2009,zhou2010}, electron-~\cite{moritz2009,ikeda2009,schmitt2011}, and un-doped cuprates~\cite{ronning1998,ronning2005,kim2006}, as well as nickelates~\cite{uchida2011} which have been in the spotlight recently for their ability to host HTSC phases~\cite{Li2019,Zeng2020}.
The high-energy kink corresponds to an abrupt renormalization of the electronic dispersion close to the nodal $(\frac{\pi}{2},\frac{\pi}{2})$ point of the Brillouin zone (BZ), usually a few hundreds of meV below the top of the valence band. 
It is connected to the seemingly unperturbed dispersion at higher binding energies around $\Gamma$ $(0,0)$ through a fast and incoherent feature: the waterfall. 
In insulating samples, the renormalized dispersion is located far in the gap~\cite{ronning1998,ronning2005,kim2006}, and is promoted to the Fermi level upon hole- or electron-doping.
These anomalies are central to the understanding of HTSC since (i) they are universal, and (ii) they renormalize the quasi-particle band opening a superconducting gap below $T_{c}$.

Many different interpretations of the waterfall have been proposed, of which the most debated ones are related to the electron-phonon coupling~\cite{Lanzara:2001nr,devereaux2004,cuk2004,yang2006,graf2008,bonvca2008}, the matrix-element effects~\cite{inosov2007,alexandrov2007,rienks2014}, and the spin-polaron scenario~\cite{grober2000,borisenko2006a,borisenko2006b,valla2007,kim2006,ronning2005,martinez1991,manousakis2007,macridin2007,wang2015}.
Phonons are promising to account for \emph{low-energy} kinks which can possibly coexist with anomalies of purely electronic origin at higher energies~\cite{graf2007}.
The latter, which are the focus of this work, appear to be better accounted for by spin-polarons, i.e. electrons heavily dressed by the antiferromagnetic spin fluctuations.
In particular, the distribution of spectral weight between $\Gamma$ and the nodal point is well reproduced by spin-polarons~\cite{wang2015,macridin2007,manousakis2007}.
The occurrence of these anomalies in undoped samples would be naturally explained since a spin-polaron can be understood as a single hole moving in an antiferromagnetic background~\cite{martinez1991}, which would then survive upon doping. 
Yet this scenario remains to be firmly established.
Indeed, systematic quantitative comparison between experiments and theoretical calculations from the undoped to the hole-doped regime is missing.
Moreover, the energy scale at which the high-energy kink appears is not properly understood, neither the related issue concerning its position in momentum space. 

In this letter, we address this problem by a combined theoretical and experimental study of the spectral function of \NaCCOC\ (Na-CCOC, with number of holes $n_h \sim x$).
Na-CCOC is a well-suited system since it is free from structural transitions upon doping, it can be synthetized both in the undoped and hole-doped regime, and from a theoretical point of view it displays a simple electronic structure which eases the construction of effective models. 
We show that the anomalies present in the undoped samples ($n_h=0$) are precursors of the ones observed in doped samples until at least $n_h=0.10$.
Our cluster dynamical mean-field theory (C-DMFT)~\cite{lichtenstein1998,lichtenstein2000,kotliar2001,maier2005} calculations are in quantitative agreement with experiment, while naturally including the spin-polaron physics on the length-scale of the cluster. 
Combined with an analysis by means of simplified effective models, we unambiguously show that the kink stems from a spin-polaron.
Most importantly, we argue that the waterfall feature may rather be understood as a crossover between two \emph{momentum} regimes: one in which correlations are mainly local, and another where spin fluctuations dominate. 
The electron-magnon coupling strongly depends on the \emph{electron momentum} and cancels in the region of local correlations.
The energy scale of the kink is related to the spin-polaron bandwidth and can be accounted for precisely both in the undoped and doped cases.
%
The survival of the spin-polaron picture at doping levels in which superconductivity is observed calls for detailed investigations of these quasi-particles in the superconducting regime.

\begin{figure}
	\includegraphics[width=0.9\linewidth]{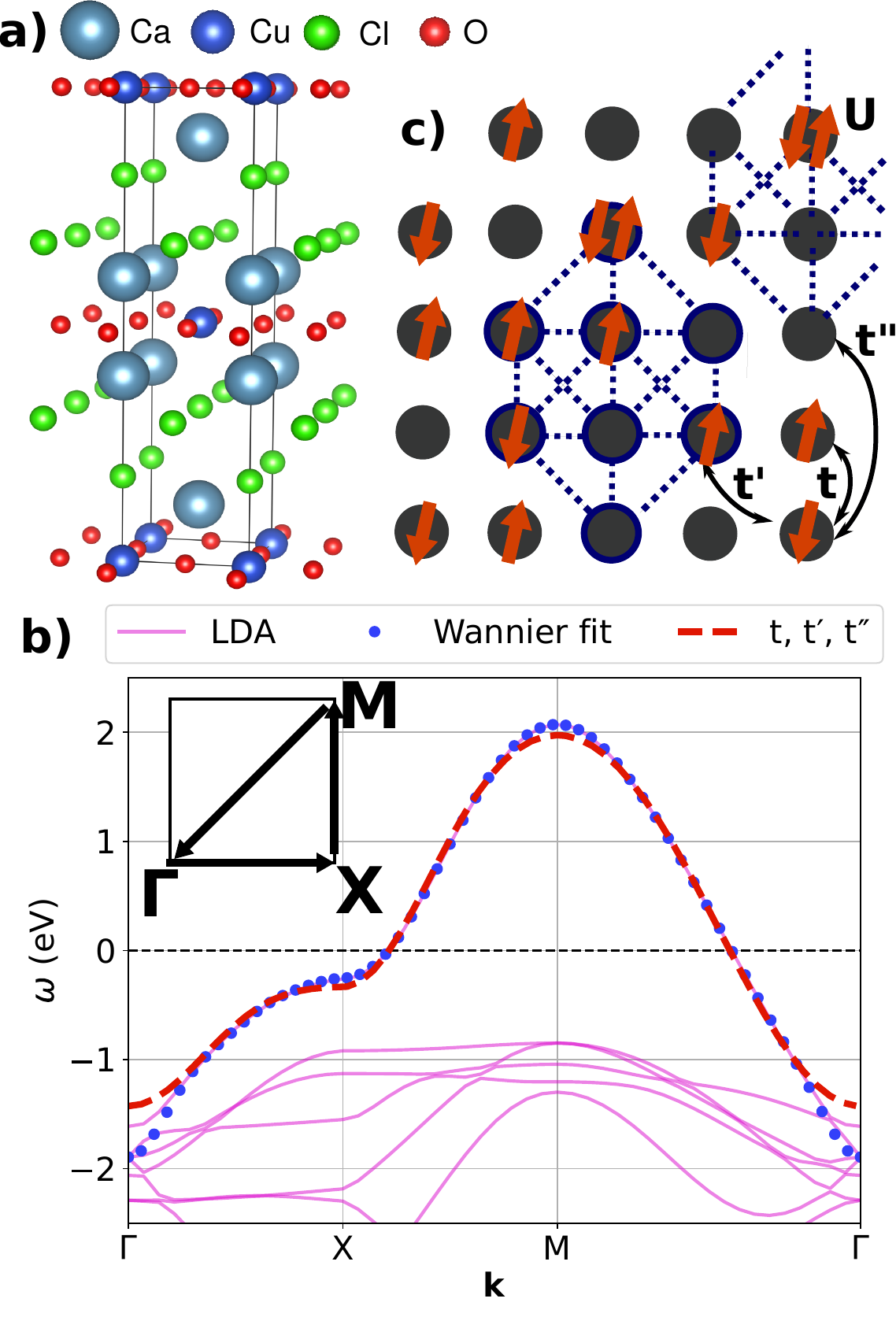}
	\caption{(a) CCOC crystal structure. 
		(b) LDA bandstructure along with the Wannier fit (blue dots), and its restriction up to the second nearest-neighbour hopping term (dashed red line). 
		The inset illustrates the BZ. 
		(c) Sketch of the single-band Hubbard model and the 8-site cluster used for C-DMFT (dashed blue lines, showing geometry and tiling).
	}
	\label{fig1} 
\end{figure}

\begin{figure*}
	\includegraphics[width=\linewidth]{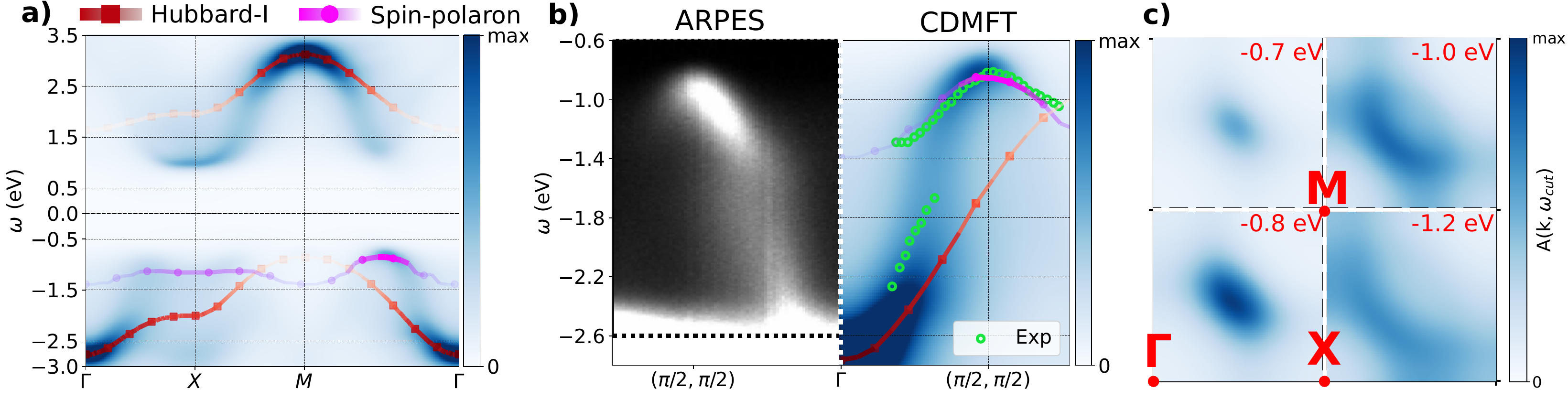}
	\caption{
		(a) C-DMFT (\emph{color plot}) \emph{vs.} Hubbard-I \emph{vs.} spin-polaron (SCBA) spectral functions for undoped CCOC.
		The latter is obtained by extracting the lowest energy dispersion. 
		(b) ARPES measurements reproduced from Ref.~\onlinecite{ronning2005} (\emph{left}) against the C-DMFT spectral function (\emph{right}). 
		The experimental dispersion is highlighted with the green dots extracted from Ref.~\onlinecite{ronning2005}.
		A rigid shift in energy was applied to the experimental dispersion, the Hubbard-I and the SCBA spectral functions (see text) to align the chemical potentials.
		(c) Calculated constant energy cuts of the upper right corner of the BZ, at four different binding energies.
	}
	\label{fig2} 
\end{figure*}

Na-CCOC single crystals were synthesized in a high-pressure cell to obtain samples with $n_h=0.06(1)$ and $n_h=0.10(1)$.
Their magnetic state was determined using a SQUID magnetometer, and their crystal quality and orientation with a 4-circle x-ray diffractometer as detailed in the supplemental information~\cite{SupMat}.
ARPES spectra were measured at the beamline BL-28 \cite{Kitamura2022} of the Photon Factory (KEK, Tsukuba, Japan) on $n_h=0.06(1)$ samples and at the Cassiop\'ee beamline of the SOLEIL synchrotron (Saint-Aubain, France) on $n_h=0.10(1)$ ones.
Single crystals oriented prior to the experiments were cleaved \textit{in situ} at low temperature and at a pressure lower than $10^{-11}$ \si{mbar}.
Photoelectron spectra were taken at photon energies of $50$ \si{eV} on both experiments. 
The temperature was kept at $T=20$ \si{K} on BL-28 for the $n_h=0.06(1)$ sample, and at $T=13 \pm 0.2$~\si{K} on Cassiop\'ee for the $n_h=0.10(1)$ one. 
The typical energy and angular resolutions were $15$ meV ($25$ meV) and $0.2^{\circ}$ ($1^{\circ}$) respectively for BL-28 (Cassiop\'ee), with a few spectra taken at Cassiop\'ee beamline with $\Delta E=12.5$ meV for better resolution around the Fermi energy.

C-DMFT~\cite{lichtenstein1998,lichtenstein2000,kotliar2001,maier2005} calculations were performed based on an effective one-band Hubbard model derived from \textit{ab initio} density functional theory calculations in the local density approximation (LDA) as described in Ref.~\onlinecite{lebert2023}, using Wien2k~\cite{blaha2020} and wannier90~\cite{mostofi2008,mostofi2014}.
Both the hopping terms of the model, $t=0.425~ \si{eV}$, $t\p/t=-0.18$, $t\pp/t=0.12$, as determined using maximally localized Wannier functions~\cite{marzari1997,Marzari2012}, see Fig.~\ref{fig1}(b,c), and the value of the local Coulomb interaction ($U/t=10.2$) as fitted from comparing to the corresponding magnon dispersion~\cite{lebert2023} are in agreement with the literature~\cite{hirayama2022}.
We emphasize that there are, thus, no hand-tuned parameters in our calculations. 
For C-DMFT, we used the continuous-time interaction expansion CT-INT solver of the ALPSCore library~\cite{shinaoka2020,gaenko2017,rubtsov2005,gull2011}, at temperatures down to $T=1/40$ eV on the 8-site cluster shown in Fig.\ref{fig1}(c).
The cluster-orientation averaged Green's function was reperiodized as described in Ref.~\onlinecite{bramberger2023}, continued analytically using the maximum entropy method~\cite{silver1990, jarrell1996} of Ref.~\onlinecite{sim2018}.
To obtain a clear physical interpretation, we compare the C-DMFT spectral function to the one obtained from the self-energy of an isolated single orbital (\emph{Hubbard-I approximation})~\cite{hubbard1963,SupMat}, as well as to the one of a hole propagating in an antiferromagnetic background (\emph{spin-polaron}) as calculated using the self-consistent Born approximation (SCBA) corrected by a three-site correlated hopping~\cite{martinez1991,wang2015}.
These two approximations can be interpreted as neglecting all non-local correlations (Hubbard-I approximation), or freezing all charge fluctuations (spin-polaron). 
We provide details in the supplemental information about how these approximations relate to C-DMFT, as well as about the crystal growth and characterization~\cite{SupMat}. 
%


In Fig.~\ref{fig2}(a) we show the C-DMFT momentum-resolved spectral function for the undoped case along $\Gamma~-X~(\pi,0)-M~(\pi,\pi)-\Gamma$. 
We observe two Hubbard bands separated by a gap of the order of $\simeq1.5~\si{eV}$, in agreement with the onset of optical conductivity at $\sim1.6~\si{eV}$~\cite{waku2004}, though slightly lower than the scanning tunneling microscopy gap of $2~\si{eV}$~\cite{ruan2016}.
The waterfall feature at the nodal point $(\pi/2,\pi/2)$ is in quantitative agreement with the ARPES measurements of Ref.~\onlinecite{ronning2005} as shown in Fig.~\ref{fig2}(b).
Not only the position in momentum matches precisely, but also the bandwidth of the renormalized dispersion located around $-1~\si{eV}$ and the distribution of spectral weight.
In Fig.~\ref{fig2}(c) we show constant energy cuts at binding energies around the top of the valence band. 
The spectral weight first increases as reaching the top of the valence band (from $-0.7~\si{eV}$ to $-0.8~\si{eV}$) and then weakens at higher binding energies, showing signs of backfolding around $-1.0~\si{eV}$.
This evolution is in excellent agreement with the remnant Fermi surface observed previously~\cite{ronning1998,hu2018}.

We have a closer look at the highly dispersive waterfall feature between $\Gamma$ and $(\pi/2,\pi/2)$, and the related kink at the binding energy $E_b\sim-1.4$ eV.
The high-intensity spectral features at $\sim-2.8~\si{eV}$ around $\Gamma$ and $\sim+2.8~\si{eV}$ around $M$ can be viewed as renormalized dispersions well captured by Hubbard-I approximation \cite{wang2015}, see Fig.~\ref{fig2}(a).
%
%
Due to the spin fluctuations, a separate quasi-particle-like feature of $\sim400$ meV bandwidth~\cite{ronning2005} emerges and leads to a kink in the spectral function, see Fig.~\ref{fig2}(b).

The phonon modes in CCOC are located at $\sim75-85$ meV \cite{lebert2020,lebert2023} which is too low in energy to account for the observed effect \cite{ronning2005}.
The fact that our C-DMFT simulations accurately capture the high-energy kink underlines its electronic origin, which can be traced back to the interaction with magnons forming a spin-polaron.
Indeed, by simulating the propagation of a hole in an antiferromagnetic spin background using SCBA \cite{martinez1991} corrected by a three-site correlated hopping term \cite{wang2015}, the quasi-particle-like dispersion is very well reproduced, see Fig.~\ref{fig2}(b)~\footnote{Note that we find the position in energy of the spin-polaron at the nodal point to be well described by SCBA when it is shifted by $U_{\mathrm{eff}}/2$, instead of $U/2$, with $U_{\mathrm{eff}}$ the screened Coulomb interaction used for the Hubbard-I approximation. It is most likely an effect of the Gutzwiller projection of the doubly-occupied states inherent to the SCBA framework, which amounts to effectively set the hopping terms to zero at half-filling, similarly to the Hubbard-I approximation. In contrast, C-DMFT on the Hubbard model takes into account the double-occupancies which are not entirely prohibited since the bandwith remains close to $U$. We provide further details in the supplemental information~\cite{SupMat}.
}.
We emphasize the crucial role of the antiferromagnetic spin fluctuations: we show in the supplemental information that single-site DMFT calculations in the insulating regime simply corresponds to the Hubbard-I results since the non-local spin correlations are neglected~\cite{SupMat}. 

The waterfall feature is hence interpreted as the crossover between a local-correlation regime and a spin-polaron band, in agreement with previous studies~\cite{grober2000}.
Yet an important question remains open: is the waterfall feature a matter of energy scales or momentum dependence?

In the electron-phonon picture the kink arises precisely at the phonon energy which entirely determines the position of the kink in momentum.
We argue that the spin-polaron picture is rather lead by the \emph{momentum} dependence of the electron-magnon coupling~\cite{SupMat}.
Indeed, in the case of a hole propagating in an antiferromagnetic background, this effective coupling reads~\cite{martinez1991}:

\begin{align}
	\mathcal{M}_{\mathbf{k,q}} = \left(\frac{1+\nu_{\mathbf{q}}}{2\nu_{\mathbf{q}}}\right)^{\frac{1}{2}}\gamma_{\mathbf{k}-\mathbf{q}}  -\mathrm{sign}(\gamma_{\mathbf{q}})\left(\frac{1-\nu_{\mathbf{q}}}{2\nu_{\mathbf{q}}}\right)^{\frac{1}{2}}\gamma_{\mathbf{k}}, 
\end{align}

where $\gamma_{\mathbf{k}}=\frac{1}{2}\left(\cos(k_x) + \cos(k_y)\right)$, $\nu_{\mathbf{q}}=\sqrt{1-(\gamma_{\mathbf{q}})^2}$, and $\mathbf{k}$ ($\mathbf{q}$) is the momentum of the hole (magnon). 
At $\Gamma$ $\mathbf{k}=(0,0)$ and at the nodal point $\mathbf{k}=(\pi/2,\pi/2)$, $\mathcal{M}_{\mathbf{k,q}}$ becomes:

\begin{align}
	\label{eq:Mkq_main}
	&\mathcal{M}_{\mathbf{(0,0),q}} = \left(\frac{1+\nu_{\mathbf{q}}}{2\nu_{\mathbf{q}}}\right)^{\frac{1}{2}}\gamma_{\mathbf{q}} -\mathrm{sign}(\gamma_{\mathbf{q}})\left(\frac{1-\nu_{\mathbf{q}}}{2\nu_{\mathbf{q}}}\right)^{\frac{1}{2}}. \\
	\label{eq:Mkq_main2}
	&\mathcal{M}_{\mathbf{(\pi/2,\pi/2),q}} = \left(\frac{1+\nu_{\mathbf{q}}}{2\nu_{\mathbf{q}}}\right)^{\frac{1}{2}}\gamma_{(\pi/2,\pi/2)-\mathbf{q}}.
\end{align}

In the local-correlation region around $\Gamma$ and $M$, the coupling vanishes because of the negative sign between the two terms in Eq.~\ref{eq:Mkq_main}, which can be interpreted as a destructive interference between two counter-propagating magnon branches.
In contrast, at $X$ and the nodal point only one term is left (see Eq.~\ref{eq:Mkq_main2}) which leads to a much stronger electron-magnon coupling. 
The C-DMFT self-energy captures these effects which lead to the strong renormalization of the dispersion at the nodal point and the characteristic incoherence of the waterfall feature.
Moreover, the kinks observed in the anti-nodal region~\cite{cuk2004} may also be related to the spin-polaron physics since the electron-magnon coupling is also strong there.
Further details, as well as an argument as of why such framework would also be relevant for the electron-doped case, are provided in the supplemental information~\cite{SupMat}.

\begin{figure}
	\includegraphics[width=\linewidth]{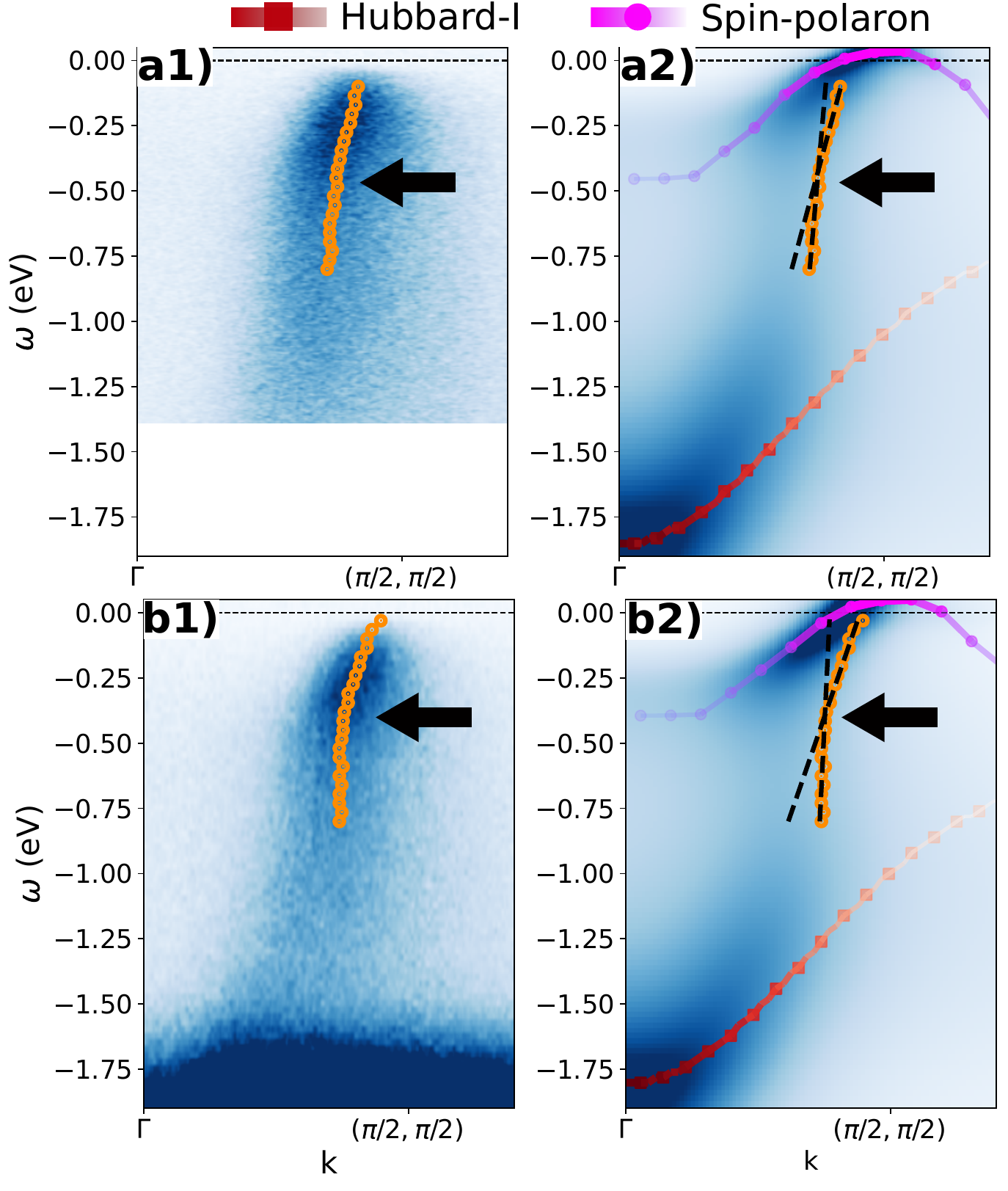}
	\caption{Experimental (\emph{left}) and theoretical (\emph{right}) quasi-particle dispersion at the nodal point $(\pi/2,\pi/2)$ for (a1,2) $n_h=0.06(1)$ ($n_h=0.05$ for C-DMFT), and (b1,2) $n_h=0.10$.
		Orange circles highlight the experimental dispersion, black arrows mark the kink position found from the intersection of two fit affine functions. 
		A rigid shift is applied to the Hubbard-I and spin-polaron spectra for alignment. 
	}
	\label{fig3} 
\end{figure}

We now turn to doped \NaCCOC\ and study the evolution of the spectral function at doping levels of $n_h=0.05,0.10$.
Figure~\ref{fig3} displays the measured and simulated spectra along the $\Gamma-M$ path showing the characteristic anomalies now promoted to the Fermi level due to the shift of the chemical potential to the top of the valence band~\cite{hu2021}.
Most interestingly, the experimental kink position around $-0.4~\si{eV}$ corresponds well to the characteristic bandwidth of the spin-polaron, which is of the same order of magnitude as the paramagnon dispersion since both are governed by the spin exchange $J$~\cite{martinez1991}.
Despite the rather large doping level of up to $n_h=0.10$, the spin-polaron scenario remains valid.
In the supplemental information, we provide an additional analysis of the spectral function and we also demonstrate that neglecting spin fluctuations at low-doping within single-site DMFT prevents the emergence of the high-energy anomalies~\cite{SupMat}. 

The survival of the spin-polaron picture is consistent with resonant inelastic x-ray scattering studies which found only little softening of the paramagnon modes of Na-CCOC upon doping \cite{lebert2020,lebert2023}, similar to other cuprates~\cite{lee2014,jia2016,betto2021}.
It is also in line with theoretical evidence from fluctuation diagnostics~\cite{gunnarsson2015,wu2017,rossi2020} and diagrammatic Monte Carlo simulations~\cite{simkovic2022a,simkovic2022b}, showing that short-range spin fluctuations, within the range of our 8-site cluster, remain strong upon hole-doping in the one-band Hubbard model. 

%
%
%
%

\begin{figure}
	\includegraphics[width=\linewidth]{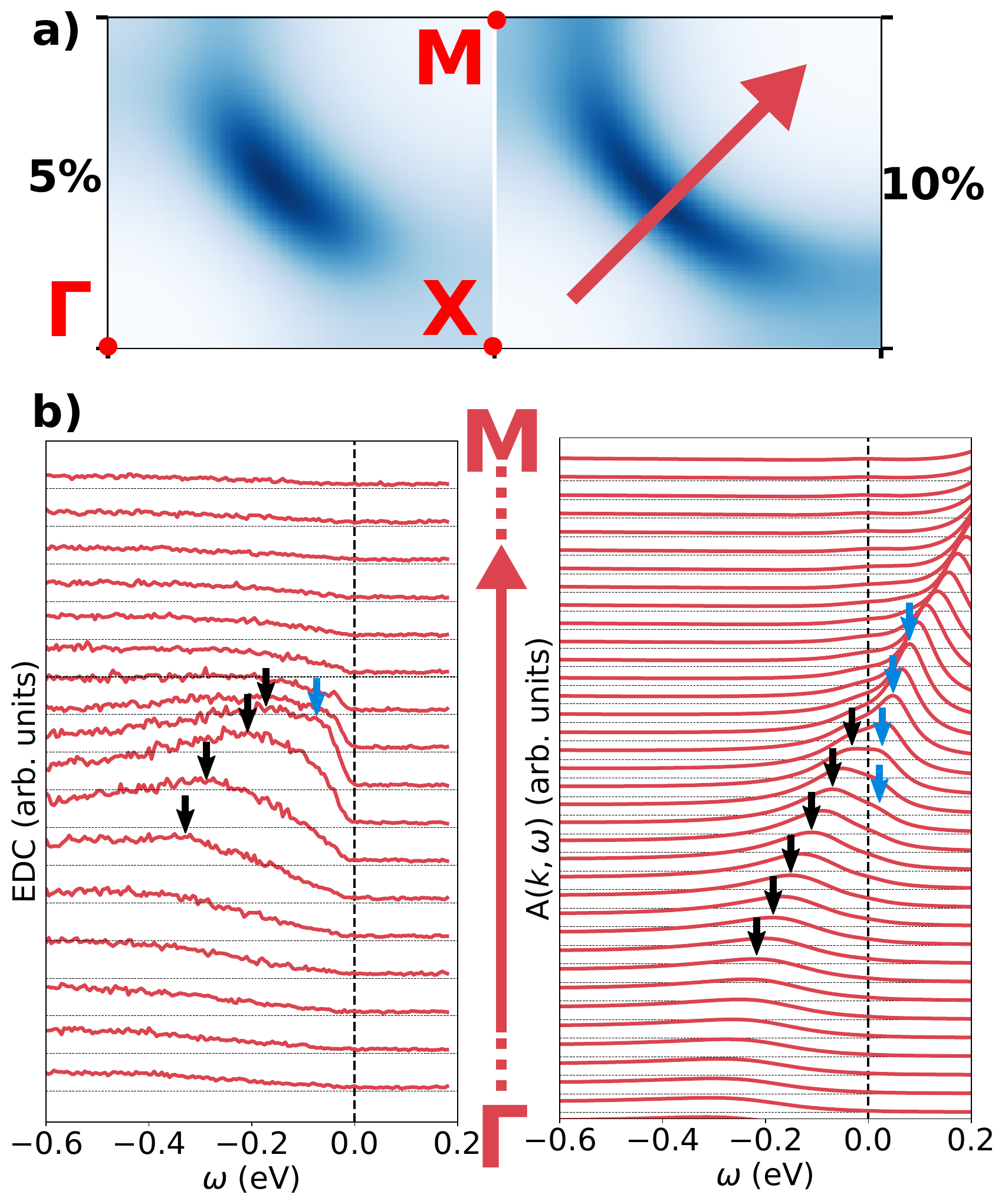}
	\caption{(a) C-DMFT Fermi surfaces for $n_h=0.05$ (\emph{left}) and $n_h=0.10$ (\emph{right}).
		(b) EDC (\emph{left}) and spectral function (\emph{right}) along $\Gamma-M$ for $n_h=0.10$.
		\label{fig4} }
\end{figure}

As for the undoped case, our observations for $n_h=0.05,0.10$ are consistent with the existing literature.
The Fermi surface evolution measured in Ref.~\onlinecite{shen2005} is correctly captured in our calculations, showing the emergence of a Fermi arc at the nodal point upon hole doping, see Fig.~\ref{fig4}(a).
%
%
We show in Fig.~\ref{fig4}(b) the ARPES energy distribution curves (EDC) and the C-DMFT spectral function.  
Close to the nodal point $(\pi/2,\pi/2)$, in agreement with Ref.~\onlinecite{kohsaka2003,shen2005}, we notice a two-peak structure in the EDCs (marked by black and blue arrows): a  well defined quasi-particle-like peak and a second 'broad hump' structure.
The C-DMFT spectrum remarkably captures this two-peak structure, and points out that the sharp feature seen experimentally is hole-like, but cut by the Fermi function.
That the measured quasi-particle peak appears further away from the Fermi level as compared to the calculations may be related to (i) an uncertainty of the effective hole-doping -as opposed to Na-doping-  in experiment, and (ii) the broadening of the quasi-particle peak well known in Na-CCOC samples~\cite{shen2005}.
The same conclusions can be drawn for $n_h=0.05$~\cite{SupMat}.


In summary, using a combined experimental and theoretical approach we firmly establish that spin-polarons are at the origin of the high-energy spectral anomalies in cuprates. 
The anomalies were, up to now, mostly discussed separately in the undoped and hole- (electron-) doped materials~\cite{zhou2010,basak2009,borisenko2006a,borisenko2006b,ronning2005,valla2007,manousakis2007,wang2015,martinez1991,macridin2007,grober2000,moritz2009,ikeda2009,schmitt2011}.
We provide a novel understanding that allows to unify the description of the high-energy anomalies at any value of doping up to $n_h=0.10$.
The anomalies appear as a crossover in \emph{momentum} between a region of essentially local correlations, and another dominated by short-range spin fluctuations, mediated by the momentum-dependent coupling between the charge carriers and magnons.
We propose a yet missing systematic comparison between experiment and a parameterized model without free parameters, at values of doping ranging from the Mott insulating regime, to the entrance in the superconducting dome, which provides a strong support to our interpretation.
Two other scenarios can be ruled out for Na-CCOC: the phonon scenario is discarded since there are no known phonon modes at the characteristic energy of the anomalies;
the matrix-elements are not sufficient either since ARPES measurements have confirmed that the high-energy anomalies were robust and consistent across multiple BZ~\cite{kim2006}. 

Further investigations of this feature at optimal doping and low-temperature are needed to map out the limits of the spin-polaron picture and shed light on its link with high-temperature superconductivity. 
In this context, our work resonates with recent efforts based on diagrammatic expansions beyond DMFT that attempt to bring together the local Mott physics along with the non-local nature of the spin fluctuations~\cite{stepanov2018a,stepanov2022,chatzieleftheriou2023}.

\section{acknowledgments}
We thank Simon Moser for enlightening discussions.
B.B.-L. acknowledges funding through the Institut Polytechnique de Paris and the Institut Quantique of Sherbrooke University. 
We are very grateful to Nathan Bujault, Felix Morineau, Carley Paulsen and Andr\'e Sulpice for help in the magnetization measurements.   
We acknowledge further help for sample synthesis at Institut NEEL from Stefan Schulte, Anne Missiaen, Murielle Legendre and C\'eline Goujon. 
M.dA would like to thank Pierre Toulemonde for advice and suggestion on the sample synthesis and characterization, as well as critical reading of this work. 
We acknowledge supercomputing time at IDRIS-GENCI Orsay (Project No. A0110901393) and we thank the CPHT computer support team.
We are very grateful to Andrés Felipe Santander-Syro for scientific and technical advise in writing  the synchrotron proposal.
We acknowledge beam time at Photon Factory and SOLEIL synchrotrons. 
%
%
Figure~1a) was generated using the VESTA software package~\cite{momma2011}. 
We acknowledge the use of the 4-circle diffractometer of ICMG-UAR2607 crystallographic platform as well as the one of IRIG, SYMMES (UMR 5819, UGA, CEA, CNRS), with the precious help by Christian Philouze and Jacques Pecaut respectively. 

\textit{Author contributions}: B.B.-L., C.F., M.dA., S.B and B.L. wrote the manuscript with input from all authors. 
The theoretical calculations were performed by B.B.-L., the analysis and interpretation of the theoretical results has been done by B.B.-L., S.B. and B.L. 
Sample synthesis, characterization and preparation was carried out by C.F., D. S.-C., H.Y., M.A., I.Y. and M.dA. 
The authors B.B.-L., C.F., H.C., Y.Ok, Y.Ob., P.L., F.B., M.dA., S.B and B.L. participated to the ARPES measurements which were analyzed by C.F. with help from  H.C, Y.Ob., B.B.-L. and M.dA.
K.H. and H.K. set and maintain BL-28 and prepared it for this experiment.
The project was conceived by B.L., M.dA. and S.B.

\bibliography{literature}
\bibliographystyle{apsrev4-2}

\renewcommand\thefigure{\thesection S\arabic{figure}}    
\setcounter{figure}{0}    
\renewcommand{\theequation}{S\arabic{equation}}
\setcounter{equation}{0}    
\clearpage
\section*{Supplemental Information}

\subsection*{Experimental details}

\subsection{Samples}
\paragraph{Synthesis.}
The $n_h=0.06(1)$ samples were obtained using a Conac press with toroidal anvils at Institut N\'eel. 
We used the following precursors: 
CaCO$_{3}$ (99.95\%), CuO (99.999\%), CaCl$_{2}$ (96\%), NaClO$_{4}$ ($\geq$ 98\%) and NaCl (99.99\%). 
First, we prepared a stoichiometric Ca$_{2}$CuO$_{2}$Cl$_{2}$ powder by a solid state reaction of CaCO$_{3}$, CuO, and CaCl$_{2}$ as described in previous works \cite{Hiroi1994, Kohsaka2002, Yamada2005}.
In an argon filled dry box, we mixed the resulting Ca$_{2}$CuO$_{2}$Cl$_{2}$ powder with NaClO$_{4}$, and NaCl precursors in a molar ratio of 1:0.2:0.2. 
We charged the mixture in cylindrical Pt capsules which were inserted in high-pressure assemblies. 
We compressed the sample at 2 GPa in order to dope with Na the Ca$_{2}$CuO$_{2}$Cl$_{2}$ precursor. 
The reacting mixtures were heated up to 1000$\degree$C at a rate of 81.5$\degree$C/min, kept at this temperature for 30 minutes and then slowly cooled down to about 870$\degree$C at a rate of 20$\degree$C/h, in order to grow crystals, and finally quenched. 
After heat treatment, we released the pressure. 
Magnetic susceptibility measurements performed using a METRONIQUE\copyright~ SQUID magnetometer on the whole batch showed no superconductivity down to $2~ \si{K}$, and were further extended on a large sample of the same batch down to $70~ \si{mK}$ using an \textit{in-house} SQUID magnetometer with dilution cryostat, giving an upper hole-content limit of $0.07$~\cite{ohishi2005}. 
Since the nominal Na substitution is difficult to assess, and also that part of the hole doping may be attributed to Ca vacancy~\cite{Yamada2005}, we decided to estimate the hole content instead.
The latter was estimated at $n_h=0.06(1)$, by comparing the lattice parameters measured by x-ray powder diffraction with synthesis of superconducting samples using the same apparatus, and for which we extrapolated the hole content from literature.
The details will be given elsewhere~\cite{NaCCOC_insitu}.
For the sample with $n_h=0.10(1)$ doping, the synthesis is described in Ref. \onlinecite{lebert2023}. 
We measured an onset critical temperature $T_c^{on}$ of 12.7(2) K for the crystal used in this experiment using an MPMS3 Quantum Design\copyright~ SQUID magnetometer. 
A ground sample from each of the two batches was analyzed with X-ray powder diffraction and Le Bail analysis of the data is in both cases compatible with the expected tetragonal space group I4/mmm, with lattice parameters of a=3.85420(3) and c=15.1151(5) for the $n_h=0.06(1)$ batch and a=3.8511(2) and c=15.1204(7) for the $n_h=0.10(1)$ one. 

\paragraph{Single crystal characterization.}
The crystalline quality of the samples was checked, and the orientation of the facets determined using two 4-circle diffractometers: a Nonius apparatus using a Incoatec micro Mo-target X-ray source equipped with Montel optics and a Bruker APEXII detector for the sample with $n_h=0.06(1)$, and an Oxford Diffraction Xcalibur S using a Mo anode source and a Sapphire CCD detector for the one with $n_h=0.10(1)$. 

\subsection{Details of the ARPES experiments} 

Propositions were made to evaluate the effective doping using the area between the Fermi arc and the antiferromagnetic BZ~\cite{meng2011}.
This method remains however approximate, and does not seem to be necessarily more accurate than relying on the trends of lattice parameters. 
A further limitation in the determination of the \emph{true} Fermi surface is the lack of gold reference sample on the BL-28 beamline during our measurements. 
Thus, we followed the procedure used in Ref.~\cite{shen2004}, i.e. the chemical potential was set at the top of the valence band (the onset of the band's spectral weight). 
This provides consistent results in comparison to $n_h=0.10(1)$ for which the Fermi level was determined from a reference gold sample at Cassiop\'ee beamline.

\subsection*{Additional spectra}

\begin{figure*}[tb]
    \includegraphics[width=\linewidth]{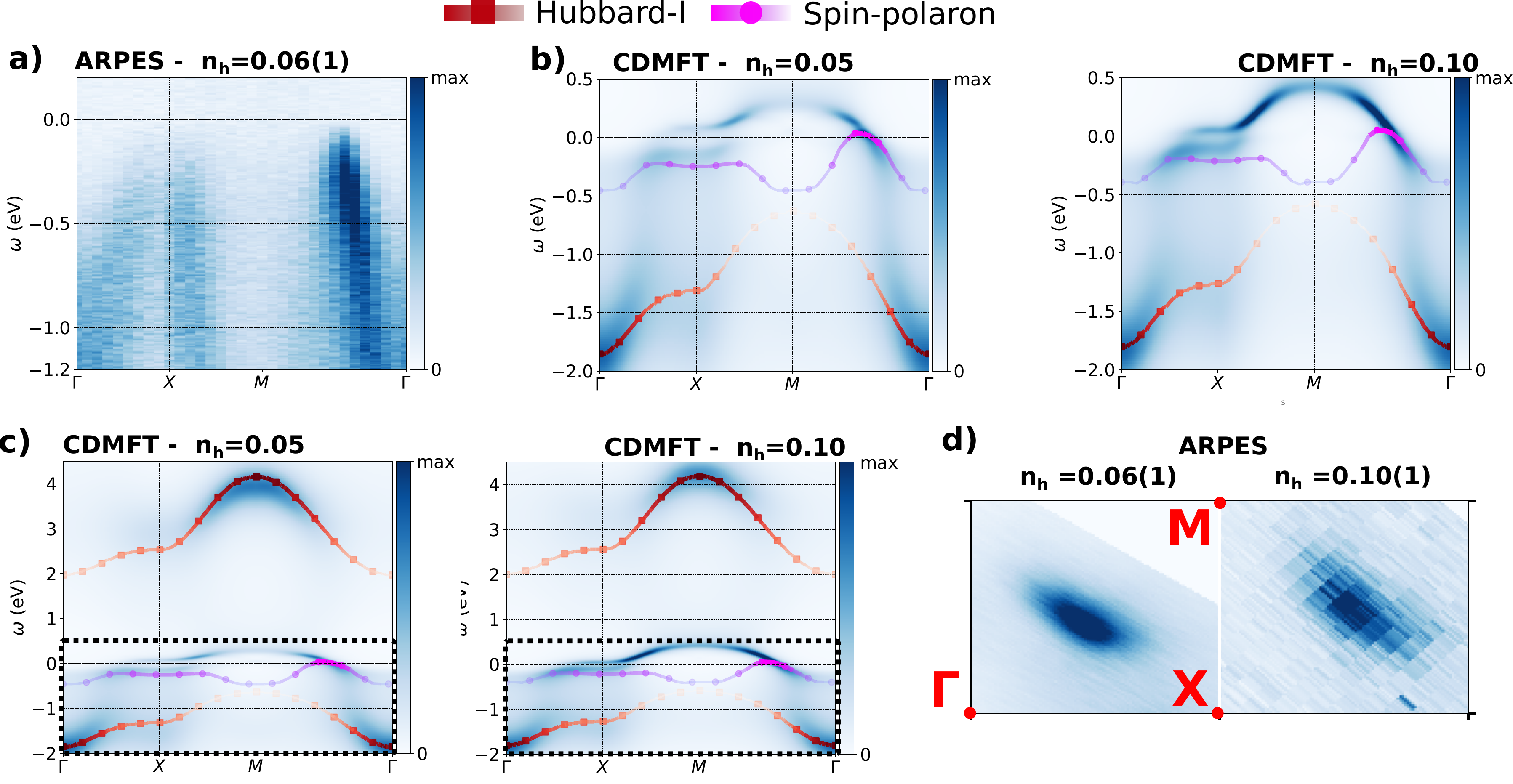}
	\caption{Spectral functions of doped Na-CCOC (a) measured on a $n_h=0.06(1)$ doped sample, and (b) computed for both $n_h=0.05$ and $n_h=0.10$.
    (c) Spectral function over the entire frequency range.
    The dotted frame denote the energy window on which we zoom in (b) for a better visualization around the Fermi level. 
    We applied a rigid energy shift to both the SCBA and Hubbard-I dispersions for alignment. 
    (d) Upper right corner of the experimental Fermi surface for $n_h=0.06(1),0.10(1)$. 
	}
    \label{fig:AddsSpec_1} 
\end{figure*}

To complement the figures presented in the main text, we provide here some additional spectral functions obtained from C-DMFT calculations for the 8-site cluster at doping $n_h=0.05,0.10$, and ARPES spectra for the $n_h=0.06(1)$ sample.  
We also present a representative example of Hubbard-I and SCBA spectra, and summarize the set of parameters used for these calculations in Table.~\ref{tab:params}.

Fig.~\ref{fig:AddsSpec_1} shows the measured spectral functions for $n_h=0.06(1)$ and a comparison to computed ones at doping $n_h=0.05,0.10$.
The waterfall feature can be seen in the ARPES data, see Fig.~\ref{fig:AddsSpec_1}(a), as well as the broad anti-nodal feature around $X$ which vanishes below the Fermi level, clearly indicating that the sample is in the pseudogap phase. 
This is also confirmed by the measured Fermi surface, as shown in Fig.~\ref{fig:AddsSpec_1}(d). 
The calculated spectra are in good qualitative agreement with the experiment, see Fig.~\ref{fig:AddsSpec_1}(b).
The maximum of spectral weight is located at the nodal point, at the Fermi energy, while the anti-nodal point shows a clear gap opening, with the two-hump feature almost symmetric around $X$. 
As discussed in the main text, the SCBA dispersion reproduces the C-DMFT features close to the Fermi energy remarkably well. 

In order to check if the momentum-driven crossover interpretation still holds at finite doping, we show in Fig.~\ref{fig:AddsSpec_1}(c) the full spectra, including both Hubbard bands. 
As for the undoped case, the Hubbard-I approximation captures the spectral weight at $\Gamma$ and $M$.
We emphasize that we used the same screened onsite interaction $U_{\mathrm{eff}}$ as for the undoped system. 
This demonstrates that the momentum-dependent electron-magnon coupling remains essential to the cuprates' physics until at least $n_h=0.10$ doping. 

Moreover, based on these results, we can infer on the evolution of the spectral function upon hole-doping. 
It is clear from our data that the introduction of holes does not lead to a simple static shift of the chemical potential. 
Instead, in the region where the electron-magnon coupling is strong, the spectral weight is transferred to the Fermi level upon doping. 
The hole-doping maintains the splitting in energy between the local-correlation region and the spin-polaron one, which was initiated already at half-filling. 
Therefore, as stated in the main text, the spin-polaron in the undoped case should be seen as a precursor of the quasi-particle present at finite doping. 

\begin{figure*}[tb]
    \includegraphics[width=\linewidth]{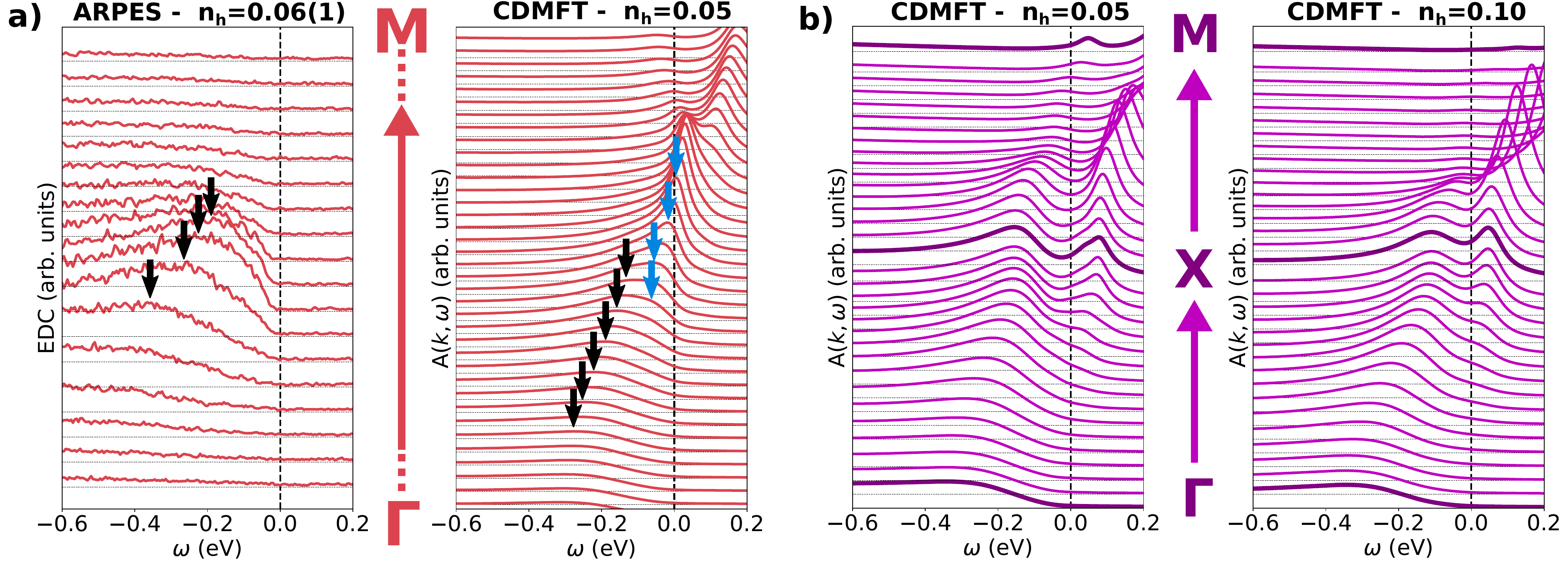}
	\caption{(a) EDC (\emph{left}) for $n_h=0.06(1)$ and C-DMFT spectral function (\emph{right}) for $n_h=0.05$ along $\Gamma-M$ . 
    (b) C-DMFT spectral function along $\Gamma-X-M$ for $n_h=0.05,0.10$.
	}
    \label{fig:AddsSpec_2} 
\end{figure*}

We show in Fig.~\ref{fig:AddsSpec_2}(a) the EDC on $n_h=0.06(1)$ and spectral function for $n_h=0.05$. 
Overall, the features resemble closely the $n_h=0.10$ data: the quasiparticle peak appears just after crossing the Fermi level, which produces a two-peak structure reminiscent of the experiment~\cite{kohsaka2003}.
As mentioned in the main text, the coherent quasi-particle peak is absent from the experimental spectrum, which is consistent with previous measurements~\cite{shen2005}.
The C-DMFT quasiparticle peak has its maximum coherence on the hole-side, as for $n_h=0.10$.
Yet, in contrast to the latter where the quasiparticle peak marked in blue stays on the hole-side of the spectrum around $\mathbf{k}=(\frac{\pi}{2},\frac{\pi}{2})$, see Fig.~4(a) (\emph{main text}), it clearly crosses the Fermi level for $n_h=0.05$.
This resembles our measured ARPES data and that of \textit{Kohsaka et al.}~\cite{kohsaka2003} who measured $n_h=0.10$ samples, for which the secondary peak crosses and appears on the electron-side of the spectrum.
These small variations between $n_h=0.05,0.10$ compounds may be related to the fact that it is difficult to reliably estimate the "true" doping of the samples, and to make a one-to-one correspondence with the theoretical doping.

The C-DMFT spectra in the anti-nodal region are displayed in Fig.~\ref{fig:AddsSpec_2}(b) for $n_h=0.05,0.10$. 
We observe the expected pseudogap around $X$, which is larger for $n_h=0.05$ since the spin-fluctuations are stronger for lower doping~\cite{Huscroft2001,civelli2005,kyung2006,sakai2009,sordi2012,macridin2006,ferrero2009,werner2009,gull2010}.
Interestingly, the dispersion shows a two-hump structure almost symmetric around $X$, which is similar to the spin-polaron dispersion shown in Fig.~\ref{fig:AddsSpec_1}.
This is consistent with the presence of a strong electron-magnon coupling in the anti-nodal region, which is detailed in the following section. 

\begin{figure*}
    \includegraphics[width=\linewidth]{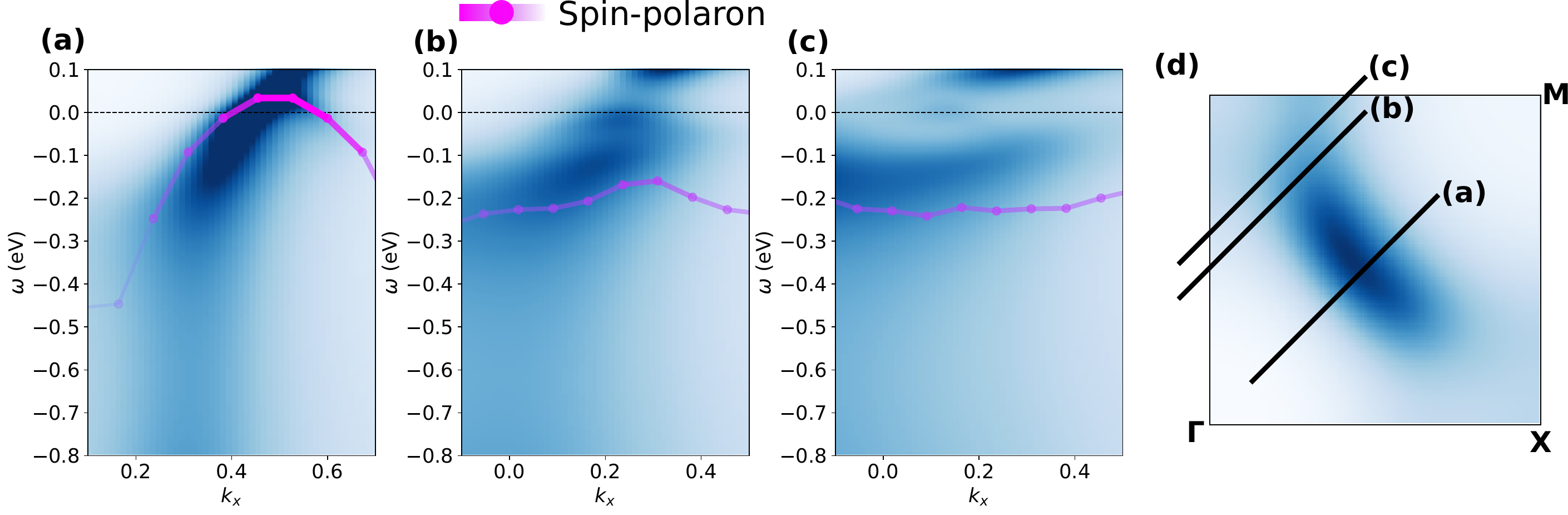}
	\caption{(a-c) C-DMFT and the SCBA spectral functions along the momentum cuts illustrated in (d), computed for $n_h=0.05$. 
	}
    \label{fig:AddsSpec_3} 
\end{figure*}

To further support our spin-polaron interpretation, we provide an additional test in which we inspect the spectral function along cuts moving away from the nodal point. 
According to the experimental data of Refs.~\cite{valla2007,zhou2010} obtained on hole-doped samples, the spin-polaron band should flatten as increasing the distance from the nodal point, and the kink should accordingly appear closer to the Fermi level.
As shown in Fig.~\ref{fig:AddsSpec_3} for $n_h=0.05$ (similar results are obtained for $n_h=0.10$), both the C-DMFT and the SCBA capture the trend correctly.
As going towards the anti-nodal region, we notice a small shift between the CDMFT and SCBA spectra, consistent with Fig.~\ref{fig:AddsSpec_1}(b).
This difference does not preclude the interpretation that the experimental decrease of the kink's energy is witnessed in our calculations by the sharp flattening of the spin-polaron dispersion, as well as the loss of spectral weight visible in both the C-DMFT and the SCBA results.  

\begin{figure}[htb]
    \includegraphics[width=\linewidth]{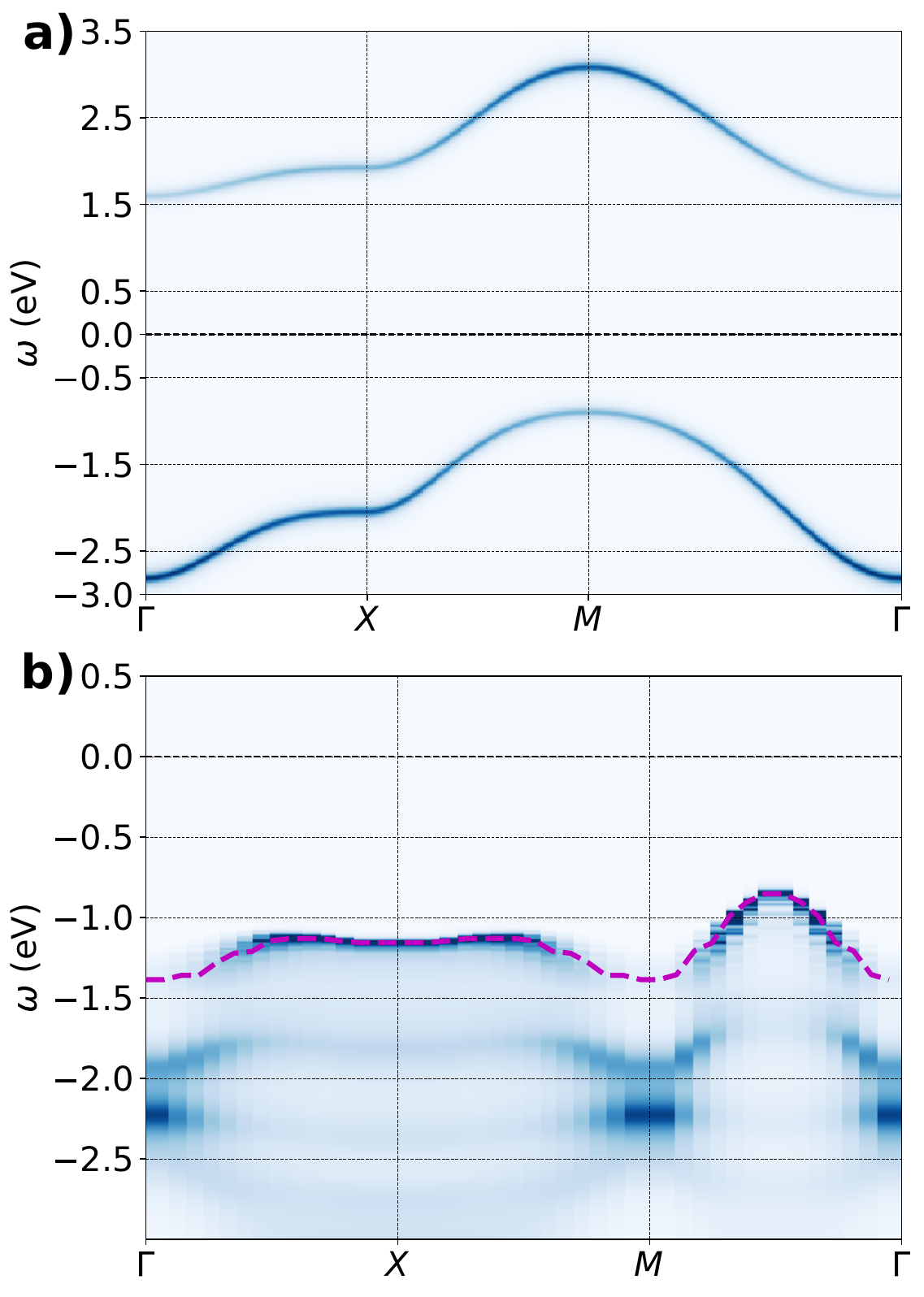}
	\caption{Representative spectral functions as obtained with (a) the Hubbard-I approximation, and (b) the SCBA, for $n_h=0$. 
    The magenta dashed line highlights the low-energy part of the spin-polaron (SCBA) spectrum which is extracted for comparison with C-DMFT. 
	}
    \label{fig:AddsSpec_4} 
\end{figure}

For completeness, we show in Fig.~\ref{fig:AddsSpec_4} a representative example of the Hubbard-I and SCBA spectral functions. 
The Hubbard-I approximation leads to two well-defined Hubbard bands, see Fig.~\ref{fig:AddsSpec_4}(a). 
The maximum of spectral weight is located around $\Gamma$ in the lower Hubbard band, and at $M$ for the upper one.
Within the Hubbard-I approximation this distribution of spectral weight is rather a matter of energy scales since it is lead by the $\frac{1}{\omega}$ divergence of the self-energy at the Fermi energy.
These maxima coincide with the coherent high-energy part of the waterfall feature as calculated with C-DMFT at $\Gamma$ and $M$.

The SCBA spectrum, shown in Fig.~\ref{fig:AddsSpec_4}(b), displays a coherent low-energy dispersion, which we extract to compare with the C-DMFT results. 
It is calculated for a $30\times30$ lattice. 
Replicas of this feature can be seen at higher binding energies.
At $\Gamma$ we notice some coherent spectral weight, located at roughly $-2.2~\si{eV}$. 
The splitting between this feature and the top of the spin-polaron band depends strongly on the hopping processes that do not disturb the antiferromagnetic order, such as the correlated three-site hopping~\cite{wang2015}, and the $t\p$, $t\pp$ terms (see next section for further details).
The splitting obtained in SCBA is lower than the one obtained in C-DMFT. 
This is a sign that corrections beyond the correlated three-site hopping are necessary for SCBA to agree even better with C-DMFT. 

Finally we summarize in Table~\ref{tab:params} the value of the hopping and (effective) interaction parameters used in the C-DMFT, SCBA and Hubbard-I calculations. 
We recall that the hopping parameters were obtained \emph{ab initio} by using maximally localized Wannier functions~\cite{marzari1997,Marzari2012}, and the local Coulomb interaction $U$ was determined by fitting the magnon dispersion measured on an undoped CCOC sample~\cite{lebert2023} with an effective Heisenberg model, whose spin exchange terms are obtained in the strong coupling limit of the Hubbard model.
Thereby, the hopping values uniquely determine the effective exchange parameters entering the SCBA equations (see next section). 
\begin{table}[h]
    \centering
    \begin{tabular}{c||c|c|c|c|c}
       Parameter & $~t~$  & $~t\p/t~$ & $~t\pp/t~$ & $~U/t~$ & $~U_{\mathrm{eff}}/t~$\\
    \hline 
       Value & $~0.425~\si{eV}~$ & $~-0.18~$ & $~0.12~$ & $~10.2~$ & $~9~$
    \end{tabular}
    \caption{Hopping terms and local interactions used in the C-DMFT, SCBA and Hubbard-I calculations.}
    \label{tab:params}
\end{table}

\subsection*{Spin-polaron picture}

The picture of the spin-polaron naturally emerges in the framework of the $t-J$ 
model, which can be derived in the limit $U\gg t$ from the Hubbard model by integrating out the doubly occupied states via a Gutzwiller projector, leading to 
the well-known $t-J$ Hamiltonian~\cite{bulaevski1968,emery1976,gros1987}:

\begin{align*}
    H_{t-J} = -t\sum_{\langle ij\rangle, \sigma}\mathrm{\tilde{c}}_{i\sigma}^{\dagger}\mathrm{\tilde{c}}_{j\sigma}^{\phantom{\dagger}} + J\sum_{\langle ij\rangle}\left(\mathbf{S}_{i}\mathbf{S}_j-\frac{1}{4}\mathrm{n}_i\mathrm{n}_j\right).
\end{align*}

Here, $\mathrm{\tilde{c}}_{i\sigma}^{\dagger}=\mathrm{c}_{i\sigma}^{\dagger}(1-n_{i-\sigma})$ are the restricted creation/annihilation operators, $\mathbf{S}_{i}$ is the spin operator on site $i$, and $J=\frac{4t^{2}}{U}$ is the spin exchange coupling. 
For the sake of simplicity, the longer ranged hopping terms $t\p$, $t\pp$ are omitted in this part of the section since they lead to hopping processes that do not disturb the antiferromagnetic background. 
We detail how they are included in practice at the end of this section.

When a hole is created, its behavior is non-trivial: it will be dressed by the spin fluctuations and form the so-called spin-polaron~\cite{martinez1991}. 
The motion of a single hole can be approximately accounted for by the self-consistent Born approximation (SCBA)~\cite{schmitt-rink1988,kane1989,martinez1991}, which we briefly summarize in the following.

The first step consists in transforming the spin operators into bosonic operators using the Holstein-Primakoff transformation for half-integer spins~\cite{holstein1940}:

\begin{align*}
\begin{split}
	&S_i^{+} = \sqrt{1-\mathrm{a}_i^{\dagger}\mathrm{a}_i^{\phantom{\dagger}}}\mathrm{a}_i^{\phantom{\dagger}}\sim \mathrm{a}_i^{\phantom{\dagger}}, \\
	& S_i^{-} = \mathrm{a}_i^{\dagger}\sqrt{1-\mathrm{a}_i^{\dagger}\mathrm{a}_i^{\phantom{\dagger}}} \sim \mathrm{a}_i^{\dagger},\\
	& S_i^{z} = \frac{1}{2} - \mathrm{a}_i^{\dagger}\mathrm{a}_i^{\phantom{\dagger}},
\end{split}
\end{align*}

where we use the linear approximation to define the bosonic operators $\mathrm{a}_i^{\dagger},\mathrm{a}_i^{\phantom{\dagger}}$.
Then, the fermionic operators can be decomposed to define a spinless hole operator:

\begin{align*}
\begin{split}
	& \mathrm{c}_{i\uparrow}^{\phantom{\dagger}}=\mathrm{h}_{i}^{\dagger}, \\
	& \mathrm{c}_{i\downarrow}^{\phantom{\dagger}}=\mathrm{h}_{i}^{\dagger}S_{i}^{+},
\end{split} 
\end{align*}

and using a Bogoliubov transformation, the $t-J$ Hamiltonian may be transformed into the spin-polaron Hamiltonian~\cite{martinez1991}:

\begin{align*}
    H_{sp} =  \frac{zt}{\sqrt{N}}\sum_{\mathbf{k,q}}\mathcal{M}_{\mathbf{k,q}}\left[\mathrm{h}_{\mathbf{k}}^{\phantom{\dagger}}\mathrm{h}_{\mathbf{k}-\mathbf{q}}^{\dagger}\alpha_{\mathbf{q}}^{\phantom{\dagger}}+h.c.\right] + \sum_{\mathbf{q}}\omega_{\mathbf{q}}\alpha^{\phantom{\dagger}}_{\mathbf{q}}\alpha^{\dagger}_{\mathbf{q}},
\end{align*}

where $z$ is the coordination number of the lattice, $N$ is the number of sites, $\omega_{\mathbf{q}}=SzJ(1-n_h)^2\nu_{\mathbf{q}}$ is the magnon dispersion with $\nu_{\mathbf{q}}=\sqrt{1-(\gamma_{\mathbf{q}})^2}$ and $\gamma_{\mathbf{k}} = \frac{1}{2}(\cos(k_x)+\cos(k_y))$. 
The magnon dispersion is damped by $(1-n_h)^2$ to take into account, in an approximate way, the hole-doping~\cite{martinez1991} ($n_h$ is set to 0 at half-filling, when only one hole exists in the lattice).
$\mathcal{M}_{\mathbf{k,q}}$ is the coupling between the hole and the magnons, and is defined as:

\begin{align}
   \mathcal{M}_{\mathbf{k,q}} = (u_{\mathbf{q}}\gamma_{\mathbf{k}-\mathbf{q}} + v_{\mathbf{q}}\gamma_{\mathbf{k}}),
   \label{eq:Mkq}
\end{align}

with $u_{\mathbf{q}}$, $v_{\mathbf{q}}$ and $\alpha^{\phantom{\dagger}}_{\mathbf{q}}$ obtained from the Bogoliubov transformation as:

\begin{align*}
\begin{split}
    & \alpha^{\phantom{\dagger}}_{\mathbf{q}} =  u_{\mathbf{q}}\mathrm{a}^{\phantom{\dagger}}_{\mathbf{q}} - v_{\mathbf{q}}\mathrm{a}^{\dagger}_{\mathbf{-q}}\\
    & u_{\mathbf{q}} = \left(\frac{1+\nu_{\mathbf{q}}}{2\nu_{\mathbf{q}}}\right)^{\frac{1}{2}}\\
    & v_{\mathbf{q}} = -\mathrm{sign}(\gamma_{\mathbf{q}})\left(\frac{1-\nu_{\mathbf{q}}}{2\nu_{\mathbf{q}}}\right)^{\frac{1}{2}}.
\end{split}
\end{align*}

\begin{figure}[tb]
    \includegraphics[width=\linewidth]{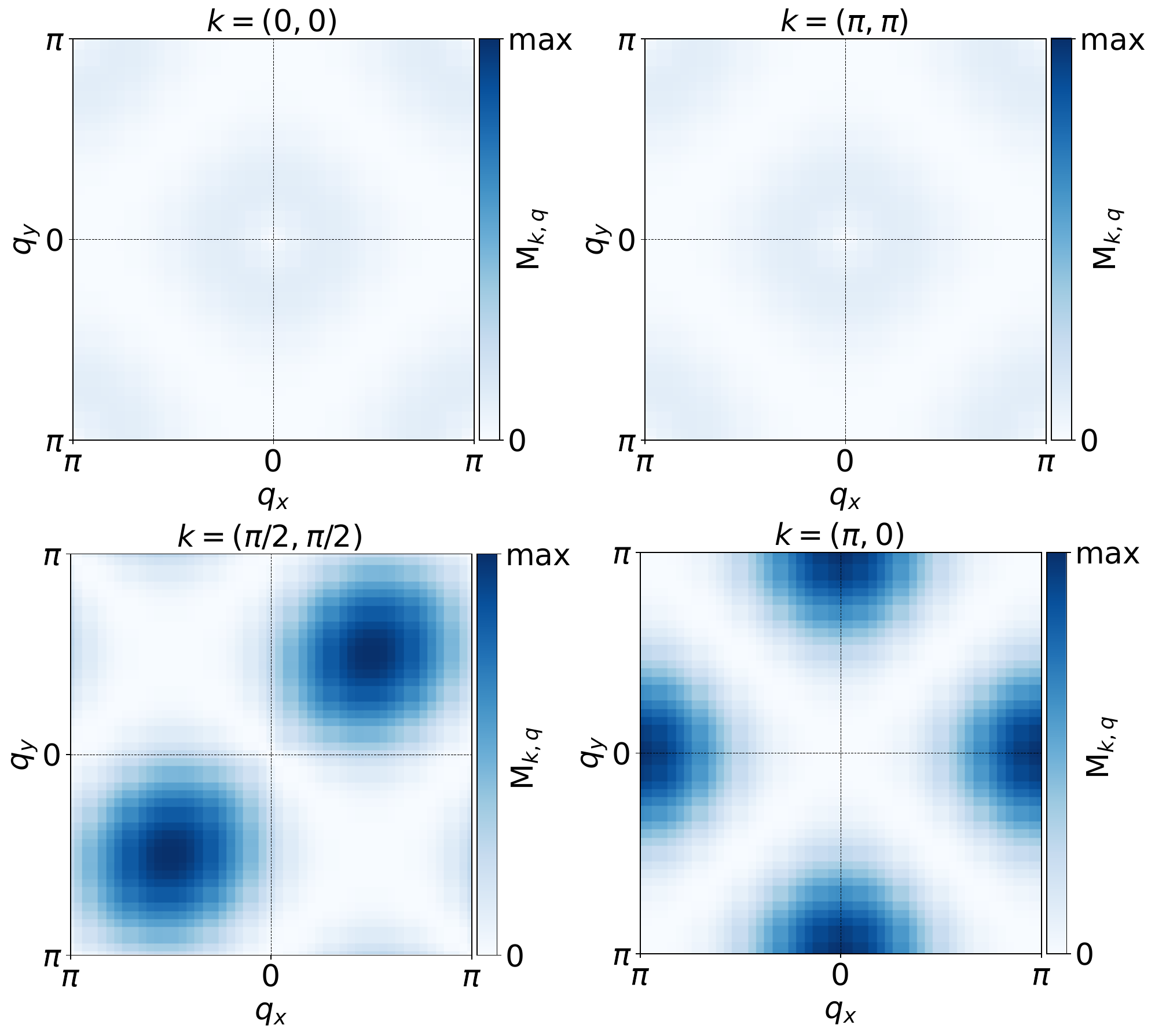}
	\caption{Absolute value of the coupling $|\mathcal{M}_{\mathbf{k,q}}|$, see Eq.~(\ref{eq:Mkq}), plotted as a function of $\mathbf{q}$ at four specific $\mathbf{k}$ points.
	}
    \label{fig:SpinPol_1} 
\end{figure}

There is no bare dispersion term in the spin-polaron Hamiltonian: the motion of the hole necessarily involves the hole-magnon coupling $\mathcal{M}_{\mathbf{k,q}}$. 
If the coupling term vanishes, the hole is localized. 
Most importantly, the coupling $\mathcal{M}_{\mathbf{k,q}}$ depends both on the momentum of the magnon $\mathbf{q}$, and on the \emph{momentum of the hole} $\mathbf{k}$.
The $\mathbf{k}$-dependence of the hole-magnon coupling is shown in Fig.~\ref{fig:SpinPol_1}, where we plot $|\mathcal{M}_{\mathbf{k,q}}|$ at fixed $\mathbf{k}$ as a function of the magnon momentum $\mathbf{q}$.
Remarkably, the coupling vanishes for all $\mathbf{q}$ at $\Gamma$ and $M$ points, i.e., exactly where the Hubbard-I provides a good description of the C-DMFT spectra. 
In the nodal and anti-nodal region, where the spin-polaron picture dominates, the coupling is maximum. 
This observation supports the view of the waterfall feature as a \emph{momentum}-driven crossover, instead of an energy-driven crossover. 

We can interpret the cancellation of the coupling in certain regions as the consequence of a negative interference between the two magnon branches $\mathrm{a}^{\phantom{\dagger}}_{\mathbf{q}}$ and $\mathrm{a}^{\dagger}_{\mathbf{-q}}$.
This is more easily seen by writing out explicitly the coupling constant as:

\begin{align*}
    \mathcal{M}_{\mathbf{k,q}} = \left(\frac{1+\nu_{\mathbf{q}}}{2\nu_{\mathbf{q}}}\right)^{\frac{1}{2}}\gamma_{\mathbf{k}-\mathbf{q}}  -\mathrm{sign}(\gamma_{\mathbf{q}})\left(\frac{1-\nu_{\mathbf{q}}}{2\nu_{\mathbf{q}}}\right)^{\frac{1}{2}}\gamma_{\mathbf{k}}.
\end{align*}

At $\Gamma$, i.e., $\mathbf{k}=(0,0)$, $\gamma_{\mathbf{k}}=1$ and $\gamma_{\mathbf{k}-\mathbf{q}}=\gamma_{\mathbf{q}}$, hence:

\begin{align*}
    \mathcal{M}_{\mathbf{(0,0),q}} = \left(\frac{1+\nu_{\mathbf{q}}}{2\nu_{\mathbf{q}}}\right)^{\frac{1}{2}}\gamma_{\mathbf{q}} -\mathrm{sign}(\gamma_{\mathbf{q}})\left(\frac{1-\nu_{\mathbf{q}}}{2\nu_{\mathbf{q}}}\right)^{\frac{1}{2}}.
\end{align*}

The minus sign signals a negative interference between the two terms.
Due to this cancellation, $\mathcal{M}_{\mathbf{(0,0),q}}$ remains small at all $\mathbf{q}$.
In contrast, at the nodal point, i.e., $\mathbf{k}=(\pi/2,\pi/2)$, $\gamma_{\mathbf{k}}=0$ and only a single magnon branch contributes to the coupling:

\begin{align*}
    \mathcal{M}_{\mathbf{(\pi/2,\pi/2),q}} = \left(\frac{1+\nu_{\mathbf{q}}}{2\nu_{\mathbf{q}}}\right)^{\frac{1}{2}}\gamma_{\mathbf{k}-\mathbf{q}}.
\end{align*}

In order to get the spin-polaron spectrum,  which we compare to C-DMFT, the self-energy is approximated with the SCBA~\cite{martinez1991}:

\begin{align}
    \Sigma(\mathbf{k},\omega) = \frac{z^2t^2}{N}\sum_{\mathbf{q}}\frac{\left|\mathcal{M}_{\mathbf{k},\mathbf{q}}\right|^2}{\omega-\omega_{\mathbf{q}}-\Sigma(\mathbf{k}-\mathbf{q},\omega-\omega_{\mathbf{q}})+i\eta}.
    \label{eq:SigmaSCBA}
\end{align}

As a consequence, the hole Green's function is strongly affected by the spin-polaron self-energy only in regions where $\mathcal{M}_{\mathbf{k,q}}$ is not vanishing.
At this stage, we can re-introduce the remaining hopping terms $t\p$, $t\pp$, as well as the correlated hopping term $J_{3s}$~\cite{wang2015}.
Since these terms do not disturb the antiferromagnetic background, they can be accounted for by adding a "non-interacting" dispersion to the SCBA self-energy:

\begin{align*}
\begin{split}
    \Sigma(\mathbf{k},\omega) = & \frac{z^2t^2}{N}\cdot\\
    &\sum_{\mathbf{q}}\frac{\left|\mathcal{M}_{\mathbf{k},\mathbf{q}}\right|^2}{\omega-\omega_{\mathbf{q}}-\epsilon_{\mathbf{k-q}}-\Sigma(\mathbf{k}-\mathbf{q},\omega-\omega_{\mathbf{q}})+i\eta},\\
    \epsilon_{\mathbf{k}} = &4t\p\cos(k_x)\cos(k_y)+2t\pp\left(\cos(2k_x)+\cos(2k_y)\right) \\
    & + \frac{J_{3s}}{2}(\cos(2k_x)+\cos(2k_y)+4\cos(k_x)\cos(k_y)).
\end{split}
\end{align*}

We also take into account the extra spin-exchange terms, including the cyclic exchange, which modify the magnon dispersion as~\cite{coldea2001}:

\begin{align*}
\begin{split}
&\omega_{\mathbf{q}} = Z_c(\mathbf{q})(1-n_h)^2\sqrt{A_{\mathbf{q}}^2-B_{\mathbf{q}}^2}\\
&A_{\mathbf{q}} = 4JS + 4J\p S(\cos(q_x)\cos(q_y)-1)\\
&+2J\pp(\cos(2q_x)+\cos(2q_y))-4J_cS^3(\cos(q_x)\cos(q_y)+1)\\
&B_{\mathbf{q}} = 2JS(\cos(q_x)+\cos(q_y))-4J_cS^3(\cos(q_x)+\cos(q_y)),
\end{split}
\end{align*}

where $Z_c(\mathbf{q})$ accounts for the quantum fluctuations and is computed following Ref.~\onlinecite{delannoy2009} (Appendix D), $J = \frac{4t^2}{U} - \frac{24t^4}{U^3}$, $J\p=\frac{4t^{'2}}{U}+\frac{4t^4}{U^3}$, $J\pp=\frac{4t^{''2}}{U}+\frac{4t^4}{U^3}$, and $J_{c}=\frac{80t^4}{U^3}$~\cite{coldea2001,delannoy2009}.

\begin{figure}[tb]
    \includegraphics[width=\linewidth]{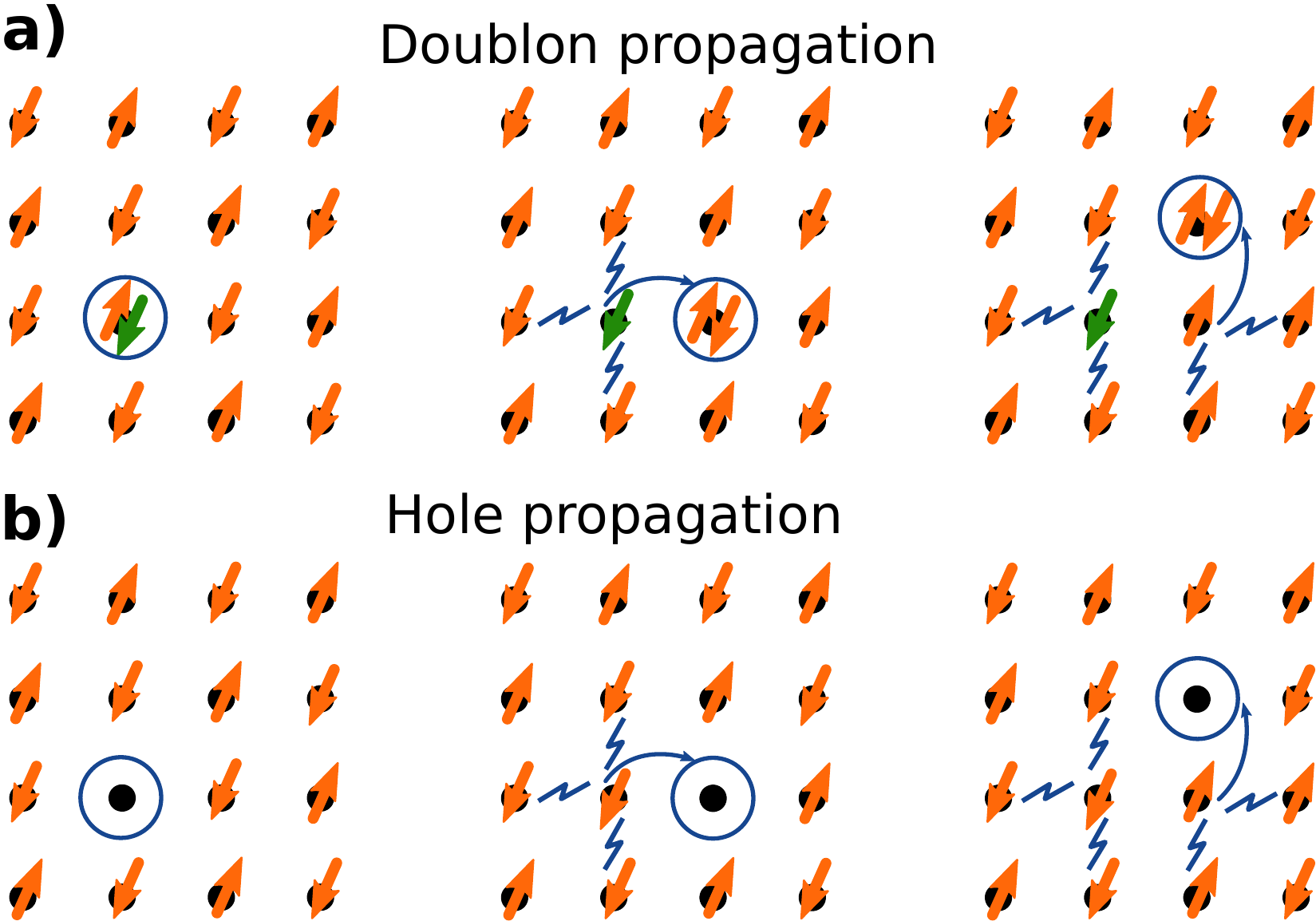}
	\caption{Sketch of the propagation of a (a) doublon, and (b) hole in an antiferromagnet. 
	}
    \label{fig:SpinPol_2} 
\end{figure}

We argue that this vision of a hole dressed by antiferromagnetic fluctuations can also be relevant for the electron-doped cuprates. 
Indeed, the antiferromagnetic order survives to large amount of electron doping in cuprates. 
Since doubly occupied sites carry no spin, they are then similar to holes from the point of view of the magnons. 
Hence the photoemission process in an electron-doped cuprate can still be approximated by the propagation of a hole in a damped antiferromagnet. 
In particular, the electron-magnon coupling would remain the same, i.e. it would remain maximum when the momentum $\mathbf{k}$ of the electron is in the nodal $(\pi/2,\pi/2)$ and anti-nodal $(\pi,0)$ regions. 
Then it is not surprising that high-energy anomalies have been observed both in the nodal and anti-nodal regions in electron-doped cuprates~\cite{ikeda2009}.

One may think of this problem in another way: starting from an antiferromagnet, what would be the behavior of an additional electron?
In other words, what would we observe in inverse photoemission?
As sketched in Fig.~\ref{fig:SpinPol_2}, the additional electron on a given site would give a \emph{doublon}. 
In analogy to the hole, the doublon carries no spin and its propagation through the lattice necessarily couples to spin excitations. 
This analogy reinforces the interpretation that spin-polarons are also relevant in the physics of electron-doped cuprates, as witnessed by the presence of high-energy anomalies in the single-particle spectrum.

\subsection*{Effective screened Hubbard interaction}

\begin{figure}[tb]
    \includegraphics[width=\linewidth]{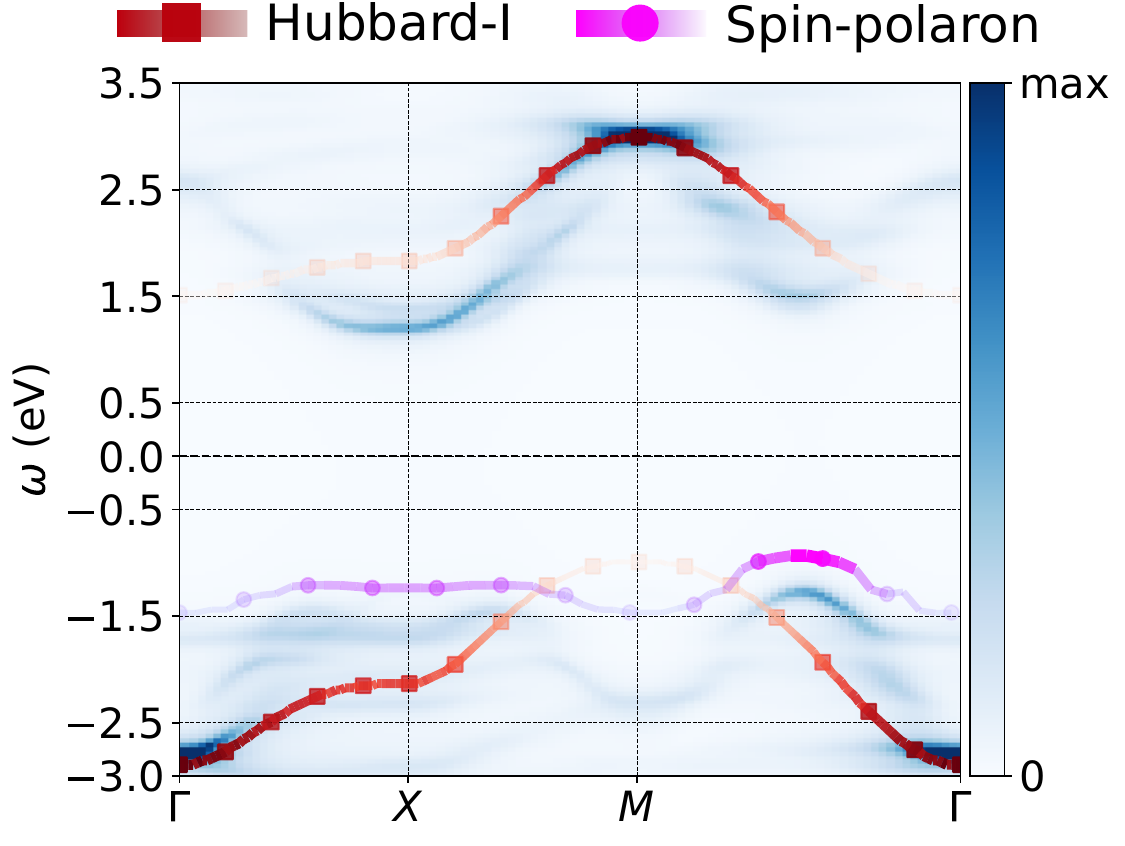}
	\caption{Spectral function obtained with an ED calculation for an isolated undoped 8-site cluster. 
    We use the same effective screened $U_{\mathrm{eff}}$ as the one used in the main text for the Hubbard-I approximation, as well as for the shift $U_{\mathrm{eff}}/2$ of the SCBA. 
	}
    \label{fig:EffU_1} 
\end{figure}

The Hubbard-I approximation reproduces the C-DMFT spectrum at $\Gamma$ and $M$, provided that it is computed with an effective screened on-site interaction $U_{\mathrm{eff}}=\simeq9t$, instead of the bare $U=10.2t$. 
The dominating screening mechanism can be understood from an exact diagonalization (ED) calculation (without bath) performed for the 8-site cluster with the PyQCM package~\cite{dionne2023}, which is shown in Fig.~\ref{fig:EffU_1}.

We first rule out that the Hubbard band splitting is not well captured by the analytic continuation, which has reduced precision away from the Fermi level.
Comparing to a calculation using ED which does not require any analytic continuation shows that it is in excellent agreement with the Hubbard-I approximation calculated using $U_{\mathrm{eff}}$. 
Moreover, we also checked that using the PoorMan's Maxent method as implemented in TRIQS~\cite{kraberger2017,kraberger2018} provides a similar result as the implementation of Ref.~\onlinecite{sim2018}. 

The screening of the on-site $U$ on the cluster via the hybridization function is also not sufficient to explain this effect.
The ED calculation, which includes no bath, shows that the hybridization has barely any effect on the splitting of the Hubbard bands. 

Therefore, the fact that the Hubbard-I approximation requires an effectively smaller $U$ is most probably the consequence of treating the hopping terms at the non-interacting level.
Indeed, the Green's function obtained from the Hubbard-I approximation reads:

\begin{align*}
    G_{H1}(\mathbf{k},\omega) = \frac{1}{\omega+i\eta+\mu-\epsilon_{\mathbf{k}}-\Sigma_{H1}(\omega)},
\end{align*}

where $\mu$ is the chemical potential, $\epsilon_{\mathbf{k}}$ is the non-interacting dispersion, and $\Sigma_{H1}$ is the Hubbard-I self-energy~\cite{hubbard1963}.
Hence the hopping terms are included in the Hubbard-I approximation, but only at the non-interacting level, in contrast to the cluster calculation which incorporates them in the correlated framework. 
In other words, the non-local correlations included in the cluster lead to an effective screening of the onsite $U$ interaction, which has to be taken into account for the local Hubbard-I approximation. 

A closer inspection of the ED spectral function brings another subject of discussion: the SCBA dispersion, which was rigidly shifted with the same energy as for the C-DMFT calculation, appears at higher binding energy than the spin-polaron feature of the isolated cluster. 
Again, analytic continuation is most probably not the culprit since using different implementations leads to very similar results.
Moreover, it is precise for the splitting of the Hubbard bands and should be even more reliable closer to the Fermi level. 
Since the hopping terms are included in the cluster both in ED and C-DMFT calculations, this shift is not due to non-local correlations within the cluster. 

\begin{figure}[tb]
    \includegraphics[width=\linewidth]{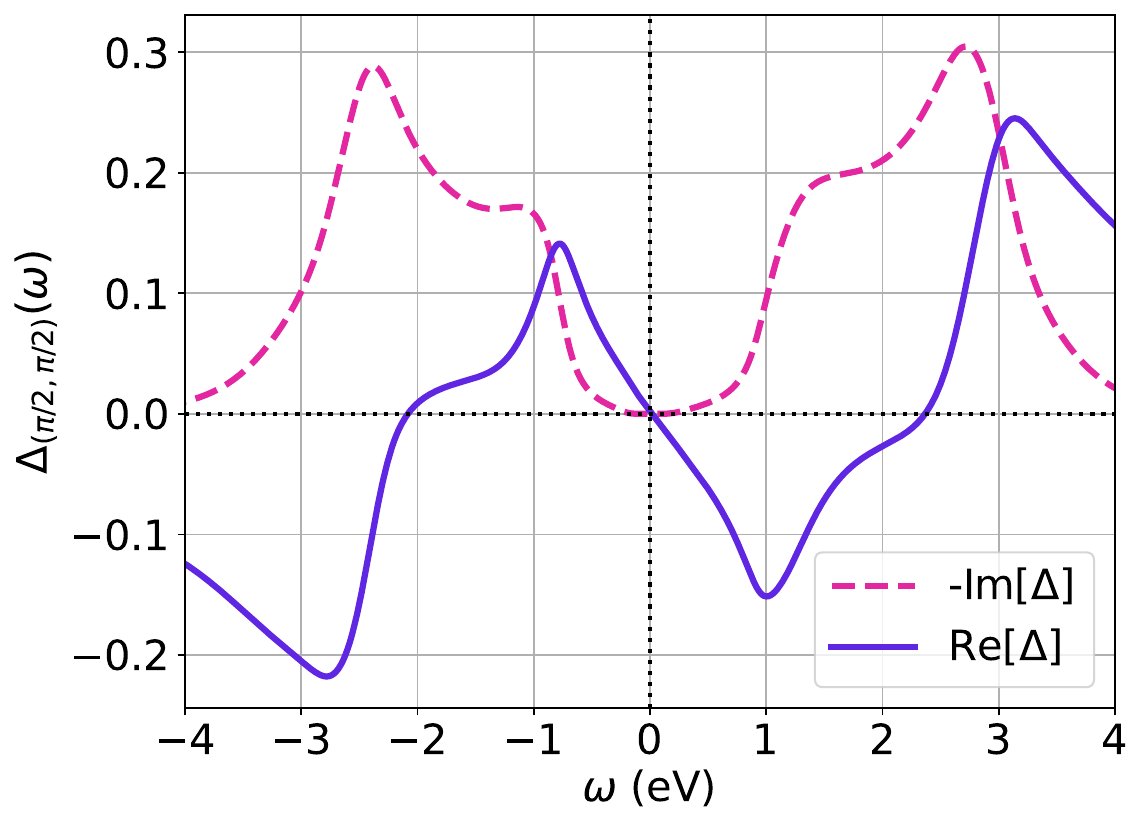}
	\caption{Real and imaginary part of the hybridization function from the C-DMFT calculation of the undoped 8-site cluster, for the cluster momentum $\mathbf{K}=(\pi/2,\pi/2)$.
	}
    \label{fig:EffU_2} 
\end{figure}

The main difference between this ED calculation compared to the C-DMFT ones is the absence of the hybridization function, which relates to the local Green's function via:
\begin{align*}
    G_{\mathrm{loc}}(\omega) = \left[\omega+i\eta+\mu-\Delta(\omega)-\Sigma(\omega)\right]^{-1},
\end{align*}
where $\Delta(\omega)$ is the hybridization function. 
We show in Fig.~\ref{fig:EffU_2} the $\mathbf{K}=(\pi/2,\pi/2)$ cluster momentum component of the hybridization function for $n_h=0$ (other components are similar).
A total of four peaks can be distinguished, two at each side of the Fermi level, at energies which correspond to the main features of the spectral function (Hubbard bands, spin-polaron). 
Most interestingly, the sign of the real part of $\Delta(\omega)$ changes betwen the "low-energy" peaks, and the high-energy ones.
Focusing on the occupied part, $\omega<0$, this sign change induces a lowering of the energy of the spin-polaron, and an increase of the energy of the Hubbard band.
In other words, the hybridization function tends to enlarge the splitting between the lower Hubbard band and the spin-polaron. 
The effect is dynamic, since including a cluster perturbation theory~\cite{Senechal2002} \emph{static} correction to the ED result does not modify the spin-polaron energy position. 
Hence the hybridization to the bath may be seen as an additional traveling channel for the spin-polaron without disturbing the local antiferromagnetic correlations inside the cluster.

Note that we need a smaller shift $U_{\mathrm{eff}}/2$ instead of $U/2$ for the spin-polaron energy position to match well with the SCBA.
This may be a consequence of the Gutzwiller projector.
Indeed, in the $t-J$ model the doubly-occupied states are projected out, whereas in C-DMFT on the Hubbard model they are taken into account and not entirely prohibited since the bandwidth remains close to the onsite $U$.
To make a connection with the screened interaction used for the Hubbard-I calculation, one may argue that the projection of the doubly-occupied states to obtain the $t-J$ model effectively amounts to setting the hopping terms to zero, at half-filling. 
Then, the same $U_{\mathrm{eff}}$ has to be used both for shifting the SCBA spectrum, and for the Hubbard-I approximation.  
This point is left for future in-depth studies about the precise relation between the SCBA and the C-DMFT spin-polaron. 

\subsection*{Comparison with single-site dynamical mean-field theory: neglecting the spin-polaron physics}

\begin{figure*}[ht]
    \centering
    \includegraphics[width=\linewidth]{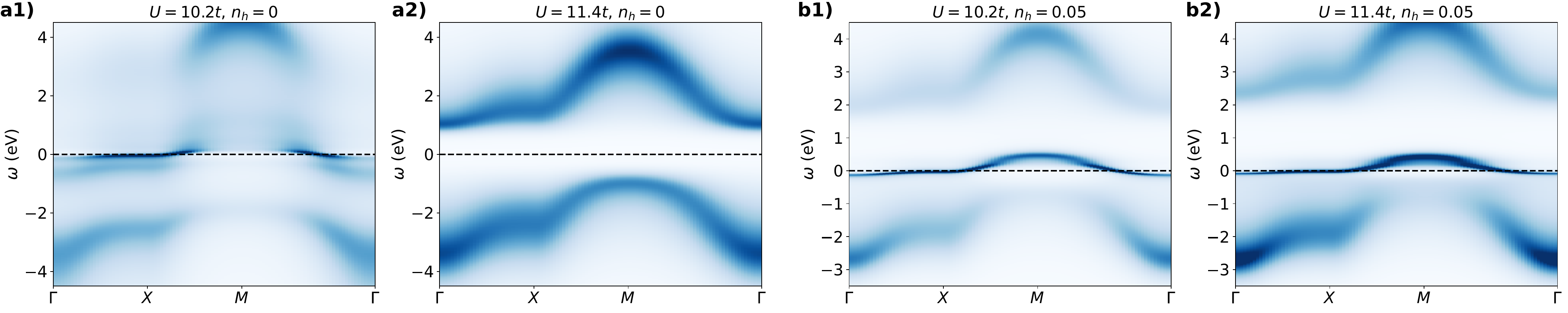}
    \caption{Momentum-resolved spectral function obtained with single-site DMFT for the undoped (a1,2) and $n_h=0.05$ hole-doped (b1,2) systems.}
    \label{fig:1s_dmft}
\end{figure*}

A recent work propose to interpret the waterfall feature as the link between the quasi-particle dispersion and the Hubbard band when a system crosses the Mott insulator-metal transition~\cite{krsnik2024}. 
In other words, spin-polarons would not be necessary to account for the high-energy anomalies which could be described using single-site DMFT. 
We argue in this section that neglecting the spin fluctuations prevents a full understanding of the high-energy anomalies. 
Indeed, single-site DMFT cannot account for the anomalies and in the insulating and low-doping regime, for which the spin-polaron physics is essential as shown in the main text. 
This conclusion goes beyond the choice of impurity solver: interaction- (CT-INT) or hybridization-expansion (CT-HYB) continuous-time quantum Monte Carlo solvers are numerically exact, as well as exact diagonalization solvers especially for single-orbital single-site impurity problems.

To support our argument, we performed single-site DMFT calculations for the undoped and $n_h=0.05$ hole-doped systems using the TRIQS library and CT-HYB impurity solver~\cite{triqs,triqs_cthyb}. 
We chose two configurations: one with the exact same $t$, $t'$, $t"$ and $U=10.2t$ parameters as in the main text, and another with a slightly larger $U=11.4t$. 
To obtain the real-frequency data we performed the analytic continuation with the TRIQS MaxEnt package~\cite{triqs_maxent}. 
The results are shown in Fig.~\ref{fig:1s_dmft}. 

First, the quasi-particle dispersion of single-site DMFT cannot explain the emergence of high-energy anomalies in undoped, insulating samples, as observed experimentally~\cite{ronning1998,ronning2005,kim2006}. 
This is directly seen in Fig.~\ref{fig:1s_dmft}(a1,2). 
When using $U=10.2t$, single-site DMFT predicts a metallic regime, hence our choice to perform the same calculation for a slightly larger $U=11.4t$ to cross the metal-insulator transition. 
As soon as the system enters the insulating phase (panels (d,h,i)), the spectrum only contains the lower and upper Hubbard bands (Fig.~\ref{fig:1s_dmft}(a2)), i.e. it basically corresponds to the Hubbard-I approximation. 
Moreover, even in the metallic regime (Fig.~\ref{fig:1s_dmft}(a1)) even if the quasi-particle dispersion emerges as a spectral weight transfer from the two Hubbard bands, it is markedly detached from them. 
This result is in stark contrast with our CDMFT calculations, which show quantitative agreement with experiment in the insulating phase. 
There, the high-energy anomalies already exist as a transformation of the Hubbard bands themselves.
We argue that they are the precursor of the renormalized quasi-particle dispersion in the hole-doped systems. 
Hence including the physics of spin-polarons is essential to unify the description of high-energy anomalies. 

Second, single-site DMFT cannot capture the waterfall at low hole-dopings neither, as shown in Fig.~\ref{fig:1s_dmft}(b1,2). 
At $n_h=0.05$, the spectrum for both $U$ values is similar to that of $Fig.~\ref{fig:1s_dmft}$(a1): well separated Hubbard-band and a quasi-particle dispersion at the Fermi level. 
In contrast to our cluster calculations, the bandwidth of the quasi-particle dispersion here is much narrower. 
Again, the quasi-particle dispersion appears as a feature detached from the Hubbard bands. 

Furthermore, the high-energy anomalies do not refer only to the waterfall, but also to the kink, i.e. the renormalization of the quasi-particle dispersion. 
At low hole-doping, there are strong experimental evidence that the Fermi surface of cuprates can display Fermi pockets~\cite{kunisada2020,kurokawa2023} that are a direct consequence of antiferromagnetic spin correlations, i.e of spin-polarons~\cite{oliviero2022,bacq2024}. 
Moreover, while the spin-polaron model reproduces well the spectrum around the anti-nodal region where the pseudogap opens, single-site DMFT cannot even account for the pseudogap physics as shown in Fig.~\ref{fig:1s_dmft}(b1,2). 
As we argue in the manuscript, the electron-magnon coupling plays an important role both in the nodal $(\pi/2,\pi/2)$ and anti-nodal $(\pi,0)$ regions, thus one should not restrict the discussion of the high-energy anomalies solely to the nodal quasi-particle dispersion.

\end{document}